\documentclass[lettersize,journal]{IEEEtran}
\usepackage{amsmath,amsfonts}
\usepackage{algorithmic}
\usepackage{algorithm}
\usepackage{array}
\usepackage{textcomp}
\usepackage{stfloats}
\usepackage{url}
\usepackage{verbatim}
\usepackage{subfigure}
\usepackage{graphicx}
\usepackage{cite}
\usepackage{color}
\usepackage{subcaption}
\usepackage{graphicx}
\usepackage{bbding}
\usepackage{algorithmic}
\usepackage{multirow}
\usepackage{booktabs}
\usepackage{float}
\usepackage{placeins}
\begin{document}

\title{MPAI-EEV: Standardization Efforts of Artificial Intelligence based End-to-End Video Coding}

\author{Chuanmin Jia, Feng Ye, Fanke Dong, Kai Lin, 
        Leonardo Chiariglione,        
        Siwei Ma,~\IEEEmembership{Senior~Member,~IEEE},\\
        Huifang Sun,~\IEEEmembership{Fellow,~IEEE},
        and Wen Gao,~\IEEEmembership{Fellow,~IEEE}
\thanks{C. Jia, F. Ye, F. Dong, K. Lin, S. Ma, and W. Gao are with Peking University, Beijing 100871, China. (e-mail: cmjia@pku.edu.cn)}
\thanks{C. Jia, S. Ma, H. Sun, and W. Gao are with Pengcheng Laboratory (PCL), Shenzhen 518055, China.}
\thanks{L. Chiariglione is with Moving Picture, Audio, and Data Coding by Artificial Intelligence (MPAI) Organization, Geneva.}
\thanks{Copyright © 2023 IEEE. Personal use of this material is permitted. However, permission to use this material for any other purposes must be obtained from the IEEE by sending an email to pubs-permissions@ieee.org.}
}

\markboth{Journal of \LaTeX\ Class Files,~Vol.~14, No.~8, August~2021}%
{Shell \MakeLowercase{\textit{et al.}}: A Sample Article Using IEEEtran.cls for IEEE Journals}


\maketitle

\begin{abstract}
The rapid advancement of artificial intelligence (AI) technology has led to the prioritization of standardizing the processing, coding, and transmission of video using neural networks. To address this priority area, the Moving Picture, Audio, and Data Coding by Artificial Intelligence (MPAI) group is developing a suite of standards called MPAI-EEV for "end-to-end optimized neural video coding." The aim of this AI-based video standard project is to compress the number of bits required to represent high-fidelity video data by utilizing data-trained neural coding technologies. This approach is not constrained by how data coding has traditionally been applied in the context of a hybrid framework. This paper presents an overview of recent and ongoing standardization efforts in this area and highlights the key technologies and design philosophy of EEV. It also provides a comparison and report on some primary efforts such as the coding efficiency of the reference model. Additionally, it discusses emerging activities such as learned Unmanned-Aerial-Vehicles (UAVs) video coding which are currently planned, under development, or in the exploration phase. With a focus on UAV video signals, this paper addresses the current status of these preliminary efforts. It also indicates development timelines, summarizes the main technical details, and provides pointers to further points of reference. The exploration experiment shows that the EEV model performs better than the state-of-the-art video coding standard H.266/VVC in terms of perceptual evaluation metric.
\end{abstract}


\section{Introduction}\label{intro}
Neural coding methods for multimedia are currently gaining more popularity, and significant efforts are being undertaken in academia and industry to explore its immanent new scientific and technological challenges. There are also significant activities in industry and standardization groups to provide artificial intelligence (AI) model-driven enablers for content production, coding, transmission, and consumption of visual media and for enhanced user experiences~\cite{basso2022ai}. AI-based video coding is crucial for providing visual entertainment, enhancing collaboration, and transforming industries in the coming years, which essentially locates at the interdisciplinary intersection between emerging technology and conventional signal processing field and has been recognized as the next battlefield of future video compression. Increasing attention from both theoretical and implementation perspectives has accelerated the evolution of visual-understanding-aware compression technology in the past five years, triggering a paradigm shift from traditional prediction-plus-transform-based hybrid frameworks to end-to-end neural approaches. As such, the demand and functional requirements of video compression need to be updated in collaboration with AI technology~\cite{ma2019image,liu2020deep,ding2021advances}. New types of video formats and content are driving advancements in video coding standards.

The classical hybrid video compression scheme (block-based predictive plus transform coding) has been established for decades, the diagram of which contains several major modules such as block split, prediction, transform and entropy coding etc. The major objective is to individually optimize the compression efficiency of each coding module. The local optimization of each module extensively promoted the rate-distortion (R-D) performances of video codecs, resulting in different families of video compression standards, such as MPEG~\cite{sikora1997mpeg,eckart1995iso}, H.26x~\cite{wiegand2003overview,sullivan2012overview,bross2021overview}, AVS~\cite{ma2015avs2,ma2022evolution,zhang2019recent}, and AOM~\cite{chen2018overview,zhao2021study} series standards. Basically, these frameworks rely on the fine-grained tuning of the rule-based methods for marginal coding efficiency enhancement. The technological progress and features of such a framework are summarized in two folds. First, the coding efficiency of such frameworks is continuously improved by incorporating plenty of crafted coding tools and expanding their candidate lists in the R-D search space. Second, hybrid coding is facing an obvious challenge, namely the performance-improving bottleneck, because the framework becomes too sophisticated to acquire higher compression efficiency within reasonable computational complexity. Different from conventional standards, MPAI (Moving Picture, Audio, and Data Coding by Artificial Intelligence) considers AI module (AIM) and its interfaces as the AI building block and AI data coding as the transformation of data from a given representation to an equivalent one more suited to a specific application.

Deep learning~\cite{lecun2015deep} has broadened the horizon of video coding. It has been demonstrated that E2E optimized image compression has even better coding performance than H.266/VVC intra~\cite{balle2018variational,cheng2020learned,chen2021end,ma2020end}. Although currently lacking sufficient evidence, neural video compression (NVC) also has enormous potential to achieve better rate-distortion (R-D) quality than existing video coding standards. Standard groups also allocate practical steps toward AI-based codec. Representative neural image and video coding activities are listed as follows. 
\begin{itemize}
    \item {\bf IEEE 1857.11} defines a set of tools for efficient image coding, including tools for encoding, decoding, and encapsulation. Some or all of the tools are composed of trained neural networks (NN) and perform block partitioning, prediction, transform, quantization, entropy coding, filtering, etc\footnote{https://sagroups.ieee.org/fvc/}. Superior R-D efficiency~\cite{chen2021end,ma2020end} has been shown that over 50\% bitrate saving can be obtained compared to HEVC/H.265 intra.
    \item {\bf JPEG AI} is working on a learning-based image coding standard offering a single-stream, compact compressed domain representation, targeting both human visualization and effective performance for image processing and computer vision tasks\footnote{https://jpeg.org/jpegai/}.
    \item {\bf JVET NNVC} denotes the Joint Video Experts Team (JVET) of ITU-T VCEG (Q6/16) and ISO/IEC MPEG (JTC 1/SC 29/WG 11) NN based video coding (NNVC), which was kicked-off in Jan. 2018 of 8-th (H-th) JVET meeting to investigate NN based technology from both efficiency and complexity perspective.
    \item {\bf MPAI} has focused its video coding activities on the MPAI Enhanced Video Coding (EVC) Evidence Project seeking to replace or improve existing tools with NNs. Another project is MPAI End-to-End Video coding (EEV) which enjoys fully NN solutions to the needs of developing an E2E video coding standard\footnote{https://mpai.community/standards/mpai-eev/}. 
\end{itemize}

Furthermore, the relationship between MPAI-EVC and MPAI-EEV is summarized as follows. To achieve intelligent video data coding, MPAI's strategy is to start from a high-performance conventional coding scheme and add AI-enabled improvements to it, with a crystal clear intellectual property (IP) policy. The functional requirements followed by a Call for Technology will be issued to cover the short-to-medium-term video coding needs within the MPAI-EVC group. Recently, the video coding research community has argued that E2E-optimized video coding schemes can realize higher performance~\cite{lu2019dvc,li2021deep,sheng2022temporal,lin2022dmvc}. However, several issues need to be examined, e.g., how such schemes can be adapted to a standard-based codec. E2E video coding promises AI-based video coding standards with significantly higher performance in the longer term. As a technical body unconstrained by IP legacy and whose mission is to provide efficient and usable data coding standards, MPAI has initiated the study of MPAI-EEV. This decision would be an answer to the needs of the many who need not only environments where academic knowledge is promoted but also a body that develops common understanding, models, and eventually standards-oriented E2E video coding.

This paper focuses on the status of standardization from MPAI-EEV with respect to coding using E2E data-trained NNs for video signals, including an overview of already adopted technologies and a summary of the larger picture of the specifications to come in this field. The paper is structured as follows. In Section~\ref{EEV-FR}, an overall systematic analysis of functional requirements and roadmap overview for the MPAI-EEV is provided and the terminology in the context of use case is set. Section~\ref{veri-model} outlines the key technology of EEV verification model. In Section~\ref{EE}, the current performance status and the experimental details are laid out. Section~\ref{discuss} provides a summary of the collaboration strategies for EEV. The open challenges and future remarks are also discussed. Section~\ref{conclude} concludes the paper.

\section{EEV: The First Attempt of AI-based end-to-end Video coding Standard}\label{EEV-FR}
Currently, the MPAI-EEV group is in the stage of functional requirement analysis for a novel video standard with AI technology. Such analysis has been recognized as the essential driving force of modern video standards, e.g., efficiency, complexity, scalability, and application scenarios. Two major issues with the highest priority in functional requirement analysis are the use case study and a corresponding verification model to verify whether such model fulfills the targeted requirements.

\subsection{Functional Requirements}
Different from conventional hybrid coding standards which are mainly designed for natural content compression, the learning-based video codecs, such as MPAI-EEV, have different technological characteristics such that they handle different application scenarios. Specifically, the neural video codecs are directly optimized for bitrate and certain quality metrics using global optimization (perpetual-oriented or fidelity-oriented). Such design simplifies the development cycle of video standards. They can also be sufficiently adaptive for the end-user requirements, for instance, visual quality by human perception and analytic-task performance. Enabled by devices with neural inference capability, these codecs could be accelerated via a massive neural parallel execution engine. The software-hardware collaboration benefits from this flexible design because they can be on-device upgraded using hot-patch via downloadable software updates. Driven by the above-mentioned features and requirements, there has been active research on neural video compression over the past five years, showing impressive rate-distortion performance and closing the gap to conventional codecs. Typical use cases of the learned codecs are listed as follows.
\begin{itemize}
    \item {\bf Long-term storage.} Videos recorded by fixed-scene cameras are usually stored for years without decoding. The requirements from such devices are a high compression ratio, random access ability, and object retrieval support\footnote{https://mpai.community/standards/mpai-evc/use-cases-and-functional-requirements/}. The computational complexity especially the decoder-end cost is not the highest priority factor.
    \item {\bf Aerial vehicles embedded camera.} Videos captured by the aerial vehicle-embedded cameras are usually not viewpoint- or scale-invariant. Specific motion characteristic and movement behavior results in difficulty in motion compensation. Motion prediction accuracy might be restricted due to large or global motion. Moreover, the distortion characteristics (bird/fish eye view) of the drone-equipped lens also make such videos distinct from natural ones.
    \item {\bf Content dependent coding.} Videos with frequent scene changes or compound textures need content-dependent coding tool sets to adapt to different metrics. Natural content optimized codec might not be appropriate enough for complex scenes. Tunable parameters directly learned from the source data and the ability to incrementally refresh the codec are of great significance when encoding such videos.
    \item {\bf Easy upgrade capability.} Fully neural video codec does not rely on a specific decoding ability such that no special-purpose hardware is required other than a general-purpose AI accelerator. This allows for reconfigurable video codec and coding tool design.
\end{itemize}

\subsection{Technical Roadmap}
Inspired by the aforementioned use cases, compression solutions with emerging technologies such as NNs have been taken into consideration for next-generation video codec study and design. The latent representations offered by NNs are also applicable for visual analysis with little modifications. The E2E optimization capability and content-adaptive learning ability naturally satisfy those requirements. More importantly, the codec parameters can be updated using the captured content to realize content-dependent coding and coding tools re-configuration.

\begin{figure*}[]
    \centering
	\includegraphics[width=0.95\textwidth]{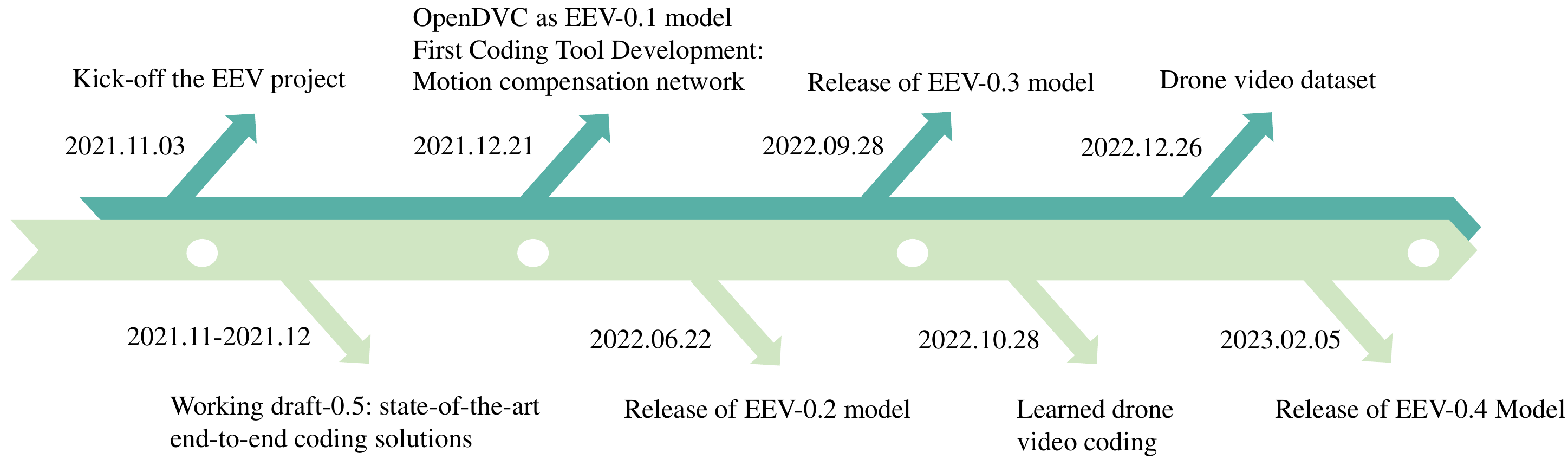}
    \caption{The technical milestones of MPAI-EEV project since its kick-off.}
    \label{fig:roadmap-eev}
\end{figure*}

To meet the demand and requirements analyzed above, MPAI-EEV project officially kicked off in Dec. 2021, studied the latest research literature, and constructed the reference model with the purpose of investigating the evidence for video codecs with the ability beyond existing standards. The ultimate goal of EEV lies in finding the near-optimal NN structure such that both video coding and network coding can be benefited. The roadmap of EEV has been depicted in Fig.~\ref{fig:roadmap-eev}. Several major milestones have been achieved from multiple perspectives. The group has released the working draft of the state-of-the-art E2E video coding solutions. Four versions of the reference model are investigated with which the exploration experiments are conducted. More importantly, the EEV aims to develop neural codecs for unmanned-aerial-vehicles (UAV) video coding and establish a public benchmark for this task~\cite{jia2023learning}.

\section{Verification Model}\label{veri-model}
The verification models of EEV have been under development in conjunction with the functional requirement discussions. As of Mar. 2023, MPAI-EEV has officially released four versions of the verification model with tags from EEV-0.1 to EEV-0.4. This section elaborates on their detailed progress and technical features of them via comparative analysis. 

\subsection{Baseline Model: EEV-0.1}
Representative pioneer research on learned video coding includes DVC~\cite{lu2019dvc} and its open-source implementation model OpenDVC~\cite{yang2020opendvc}. Based on comprehensive consideration of development feasibility, model transparency, and the basic framework license of MPAI, the OpenDVC has been adopted as the starting point of EEV verification model with agreements from all EEV experts. The authors of~\cite{yang2020opendvc} also expressed their permission and authorization for the usage of it tagged with EEV-0.1 for this project. 

\begin{figure}[]
\center
\includegraphics[width=0.48\textwidth]{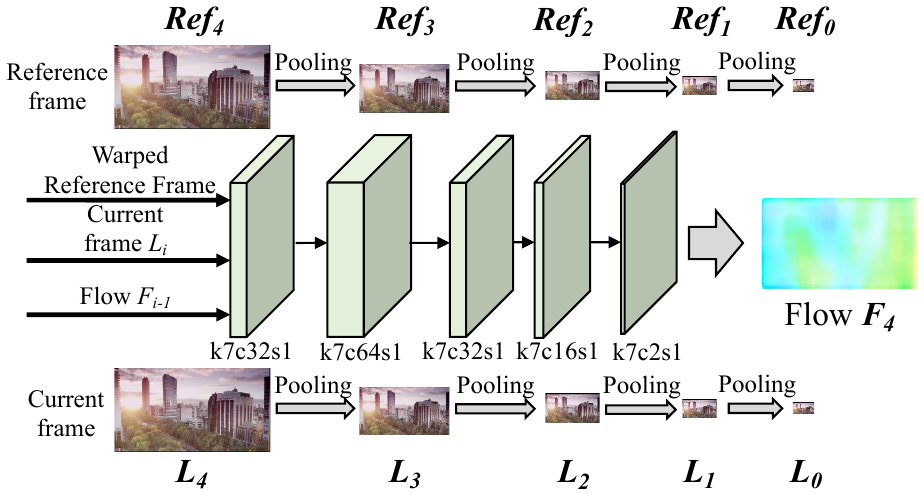}
\footnotesize
\caption{The diagram of motion estimation (ME) network adopted in EEV model. The multi-scale optical flow pyramid structure has four different pyramids, each of which is represented by the subscript in this figure. The notation ``k7c32s1'' denotes the convolution layer has 32 channels with $7\times7$ kernel and the stride is 1 pixel. Note that the reference frame here indicates $\hat{x}_{t-1}$ while the current frame corresponds to $x_{t}$.}
\label{Fig:opt-flow}
\end{figure}

\begin{figure*}
    \centering
	\subfigure[EEV-0.1]{
	\includegraphics[width=0.21\textwidth]{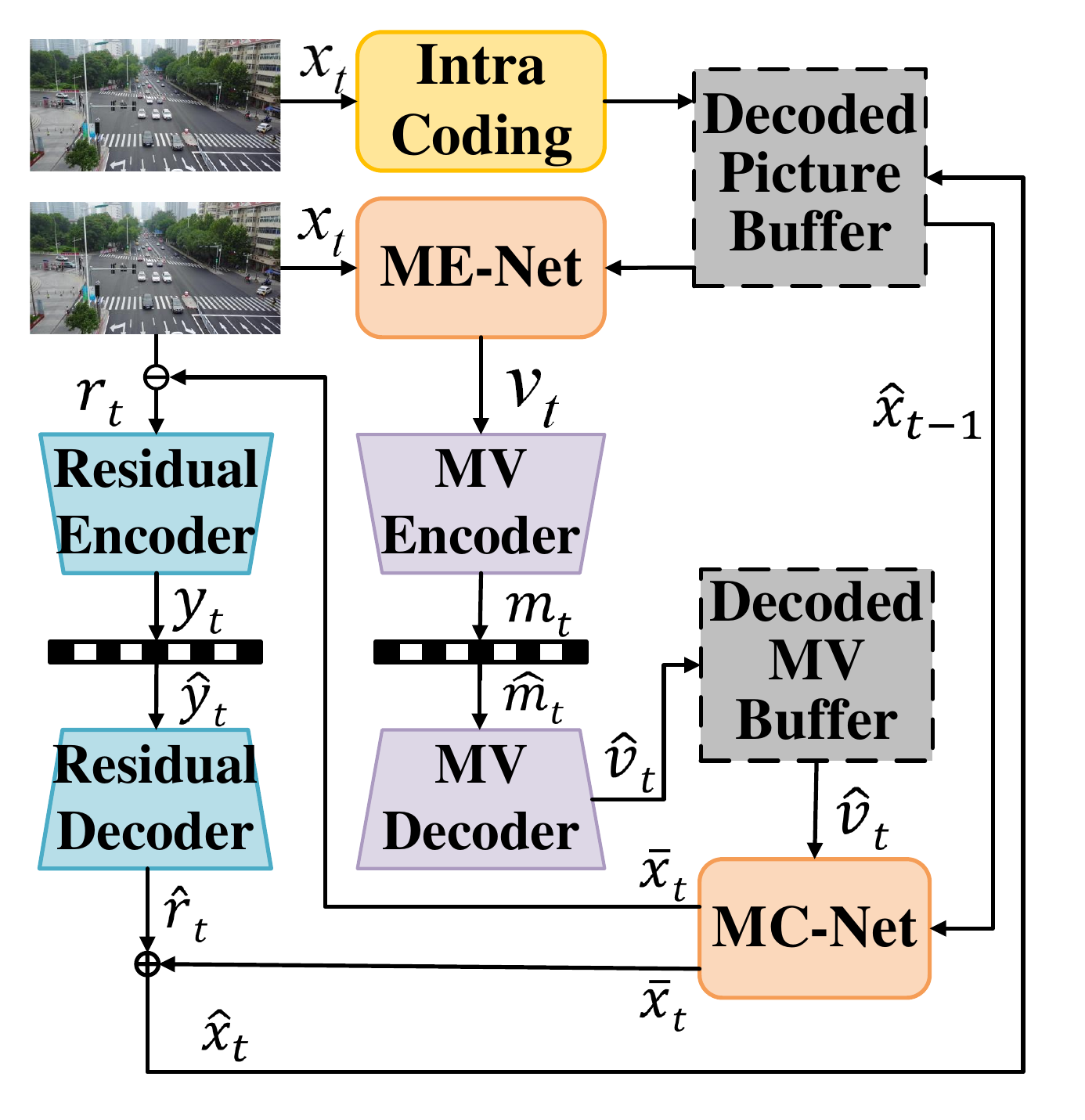}}
	\subfigure[EEV-0.2]{
	\includegraphics[width=0.2\textwidth]{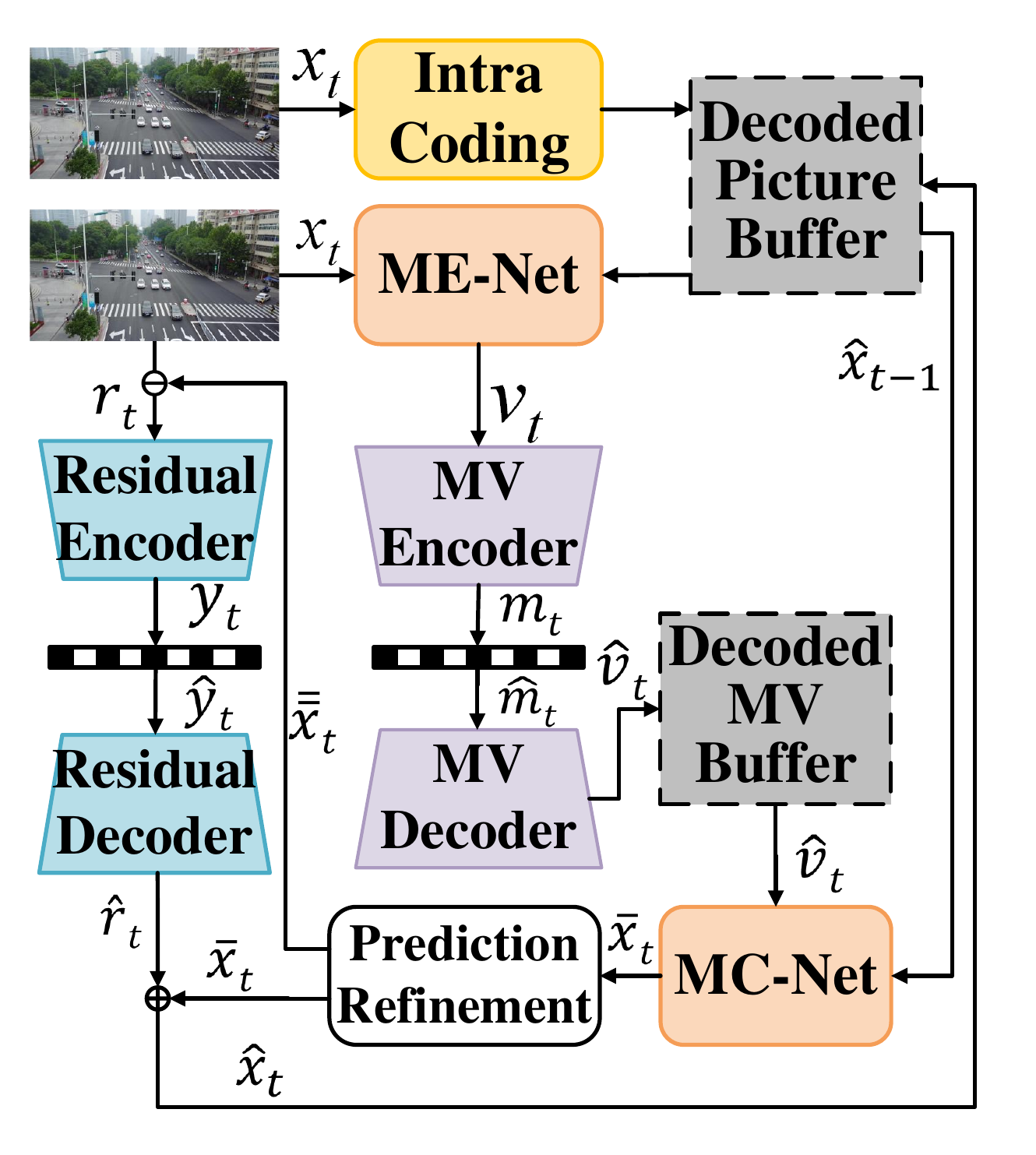}}
	\subfigure[EEV-0.3]{
	\includegraphics[width=0.25\textwidth]{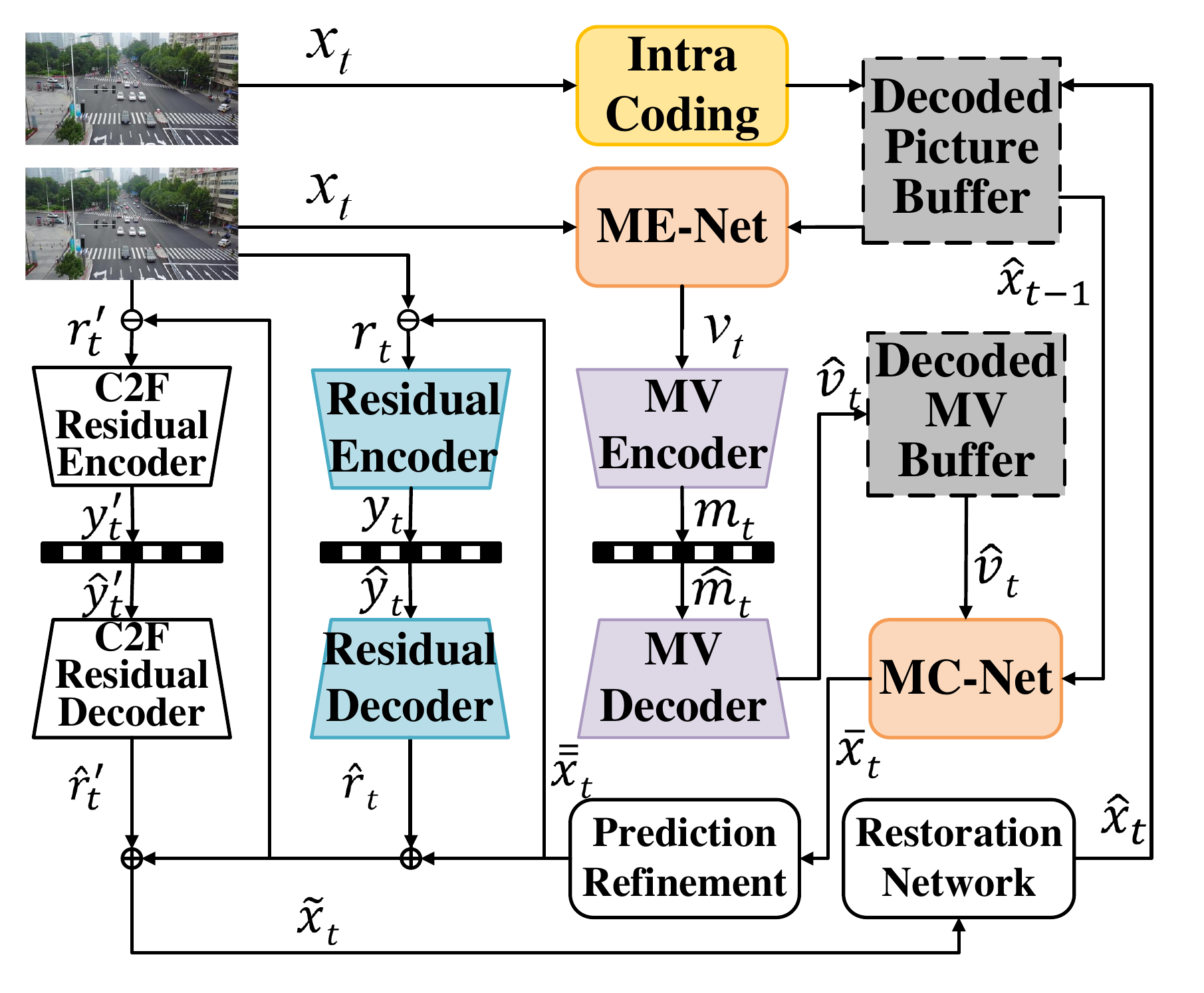}}
	\subfigure[EEV-0.4]{
	\includegraphics[width=0.24\textwidth]{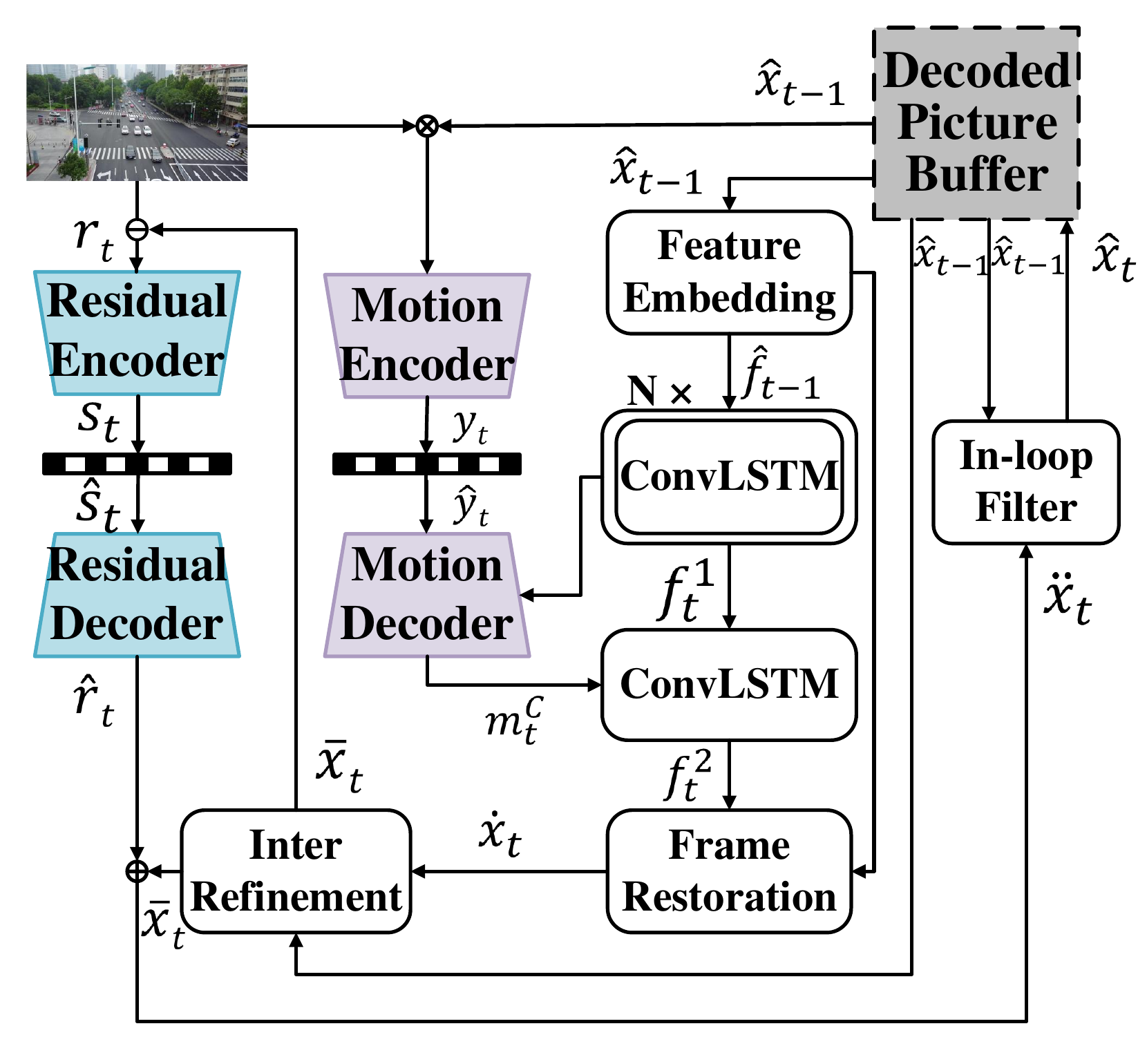}}
\caption{The block diagram of the EEV verification models EEV-0.1, EEV-0.2, EEV-0.3 and EEV-0.4. White boxes indicate the newly adopted coding methods. The EEV-0.1 has the identical framework with OpenDVC~\cite{yang2020opendvc}. EEV-0.2 and EEV-0.3 contain more coding tools on top of the starting point. EEV-0.4 introduces an advanced motion representation scheme named motion decoupling module such that the framework is relatively more complicated than its predecessors.}
\label{fig:diagram-compare}
\end{figure*}

\subsection{Enhanced method: EEV-0.2 and EEV-0.3}
With the baseline model determined, the group further studied enhanced coding tools to increase R-D efficiency via introducing optimized sub-modules to form the EEV-0.2 and EEV-0.3. The direct comparison of EEV-0.1, EEV-0.2, and EEV-0.3 are shown in Fig.~\ref{fig:diagram-compare}. Given the schematic diagram, the EEV-0.2 model could be regarded as EEV-0.1 with the enhanced motion compensation prediction (MCP) network. Specifically, existing investigations on neural reference picture enhancement carried out the conclusion that the inter-predictive coding efficiency can be significantly improved by introducing a prediction refinement network. The principle behind this design is to improve the quality of the predicted signal with highly accurate motion compensation as well as suppressing the prediction error caused by motion blur. EEV-0.3 contains two additional coding tools on top of EEV-0.2: coarse-to-fine residual modeling (C2F) and in-loop restoration network. The C2F residual modeling module establishes a multi-layer residual modeling module in a coarse-to-fine manner, allowing fine-grained textural information to be reconstructed in a scalable manner using residual coding. This module enhances the quality of the reconstructed pictures and reference pictures. The technological descriptions are presented as follows.

{\bf Inter Predictive Refinement Network.} 
To realize higher predictive coding efficiency, one key research direction is to promote the quality of the reference frames~\cite{zhao2018enhanced,zhao2019enhanced}. For each inter-coded frame $x_{t}$, the most recent compressed image $\hat{x}_{t-1}$ from the decoded picture buffer (DPB) is utilized as the reference frame. The motion estimation (ME) process is subsequently realized from $\hat{x}_{t-1}$ to $x_{t}$ by ME-Net using a multi-scale optical flow pyramid network. Shown in Fig.~\ref{Fig:opt-flow}, the ME-Net could model the translational displacement between adjacent pictures via the pixel-wise motion vector (MV). Specifically, the motion between the reference frame and the current pristine image is calculated. Following EEV-0.1, both EEV-0.2 and EEV-0.3 down-sample the picture using $2\times2$ global average pooling to realize a 5-layered multi-scale motion pyramid and learn the optical flow in a scalable fashion. For each layer, the MV is defined as follows.
\begin{equation}
    F_{i} = \mathcal{R}(Warp(Ref_{i-1}, F_{i-1}), L_{i}, F_{i-1}), 
\end{equation}
where $F_{i}$ is the optical flow of each layer $i$, $Ref_{i-1}$ is the reference frame, $Warp(\cdot)$ represents the bilinear warp operation and $L_{i}$ indicates the current frame in the motion pyramid. $\mathcal{R}$ encapsulates all learnable parameters in Fig.~\ref{Fig:opt-flow}. Thus loss function can be formulated as.
\begin{equation}
    \mathcal{L} = \Sigma_{i=0}^{4}||Warp(Ref_{i}, F_{i})-L_{i}||^{2}_{2}.
\end{equation}
Note that $F_{4}$ is initiated as $zero$ MV and $F_{0}$ is our target MV. In our implementation, the ME-Net is separately trained first and the converged weights are utilized as initializer of the ME-Net in E2E training. Since $F_{0}$ denotes the motion field from compressed image $\hat{x}_{t-1}$ to uncompressed current frame $x_{t}$, it should be encoded and transmitted to ensure the consistency between encoder and decoder.

\begin{figure}
\includegraphics[width=0.48\textwidth]{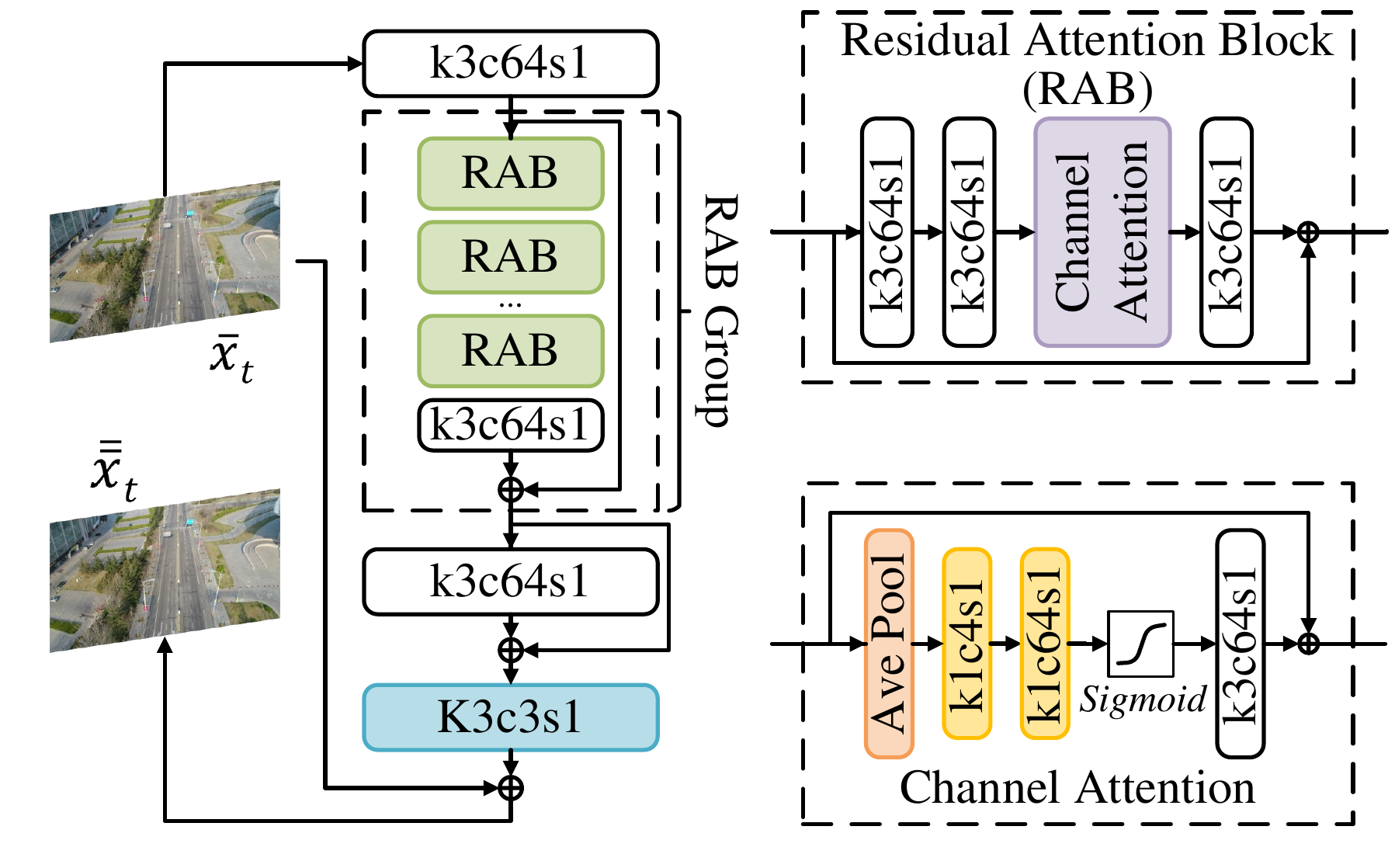}
\caption{The detailed structure of inter prediction refinement network. The convolution kernel size, number of channels, and strides are shown. There are 5 RABs in the EEV-0.3 model.}
\label{Fig:pred-refine}
\end{figure}

The prediction refinement network leverages the residual channel attention mechanism~\cite{zhang2018image}. Such structure benefits translational and non-translational motion modeling and obtains better prediction quality. The detailed structure of inter prediction refinement network is shown in Fig.~\ref{Fig:pred-refine}. During training, the prediction refinement loss $\textit{L}_{pred}(x_{t}, \bar{\bar{x}}_{t}) = D(x_{t}, \bar{\bar{x}}_{t})$ is used as part of the distortion in the learning objective, where $\bar{\bar{x}}_{t} = F_{pred}(\bar{x}_{t})$. And $F_{pred}$ denotes the inter-prediction refinement net. With the help of prediction refinement, better prediction quality could be obtained. And if the prediction frame has less difference from the original frame, the energy of the prediction residual as well as its corresponding bit-rate will be smaller. 

{\bf C2F Residual Modeling.} A simple but effective module is designed in EEV-0.3 to model the prediction residual coding in a C2F manner by cascading a second-stage residual compression. The C2F residual module has the same network structure as the previous residual stage (denoted as $\hat{r}_{t}$).
\begin{equation}
r'_{t} = x_{t} - (\bar{\bar{x}}_{t} + \hat{r}_{t}).
\label{c2f-residual}
\end{equation}
Finally, after two stages of residual compression, the compressed frame is reconstructed as follows:
\begin{equation}
\tilde{x}_{t} = \bar{\bar{x}}_{t} + \hat{r}_{t} + \hat{r}'_{t}.
\label{c2f-residual}
\end{equation}

\begin{figure}
\centering
\includegraphics[width=0.48\textwidth]{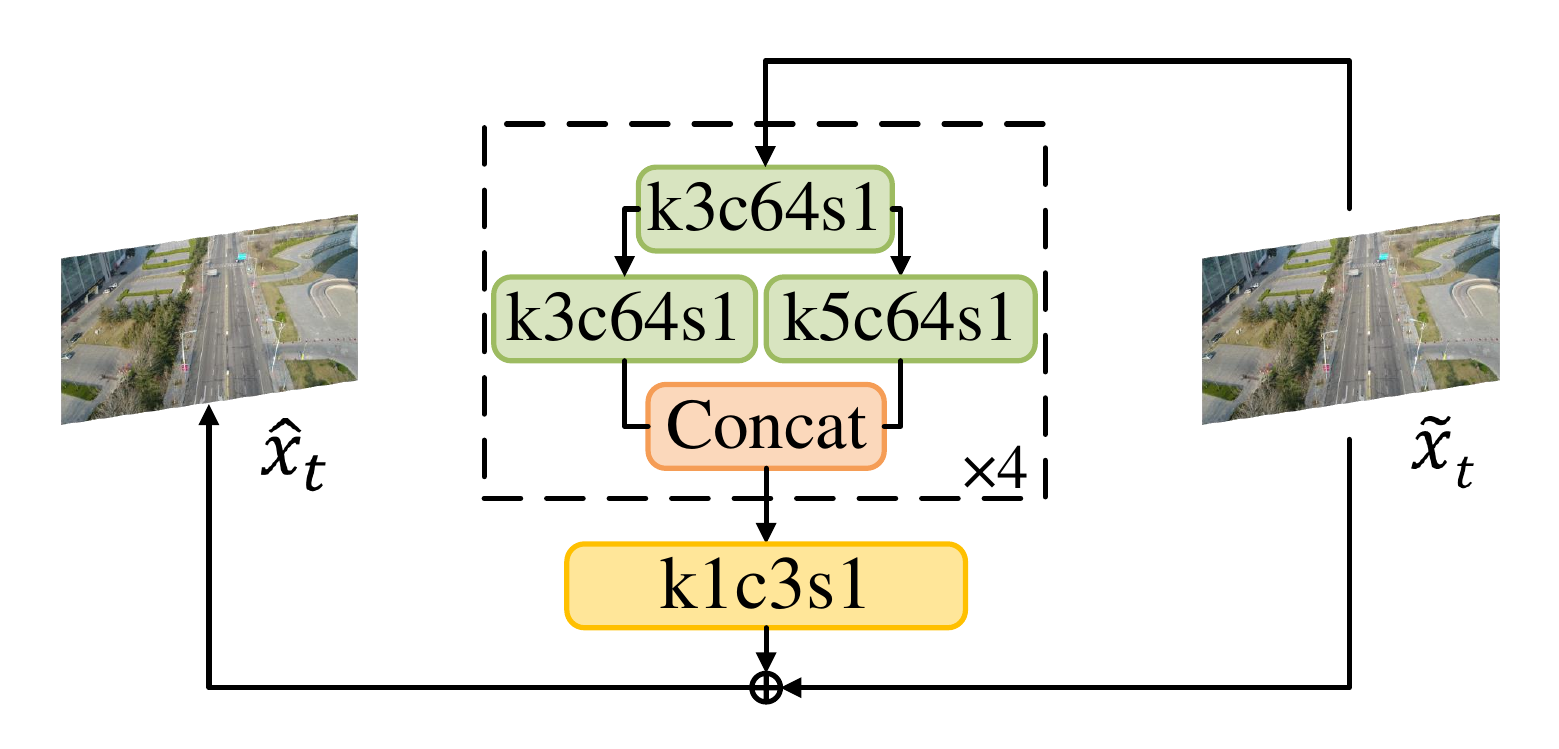}
\caption{The ILR network structure of EEV-0.3 model. Variable convolution kernels and skip shortcuts are employed. The global input-output direction residual is able to accelerate the training processing in capturing the difference between the uncompressed and the coded images. Similar notations are used with Fig.~\ref{Fig:pred-refine}.}
\label{Fig:loop-filter}
\end{figure}

{\bf In-loop Restoration Network.}
From the aspect of reducing compression noise, EEV-0.3 adopts a novel coding tool, in-loop restoration (ILR) network. Similar approaches have been shown to be effective in conventional framework~\cite{liu2020deep,ma2019image}. As shown in Fig.~\ref{Fig:loop-filter}, the in-loop restoration network is fully convolutional and has variable kernel size plus global residual connection. Following the existing in-loop restoration network structure in deep network-based conventional codecs, such design is a reasonable trade-off between performance and complexity. The input of the in-loop restoration network is the reconstruction image calculated by Eqn.~\ref{c2f-residual} and this network generates the final restored image $\hat{x}_{t}$.
\begin{equation}
\hat{x}_{t} = \mathcal{F}_{ILR}(\tilde{x}_{t}|\theta),
\label{in-loop-filter-net}
\end{equation}
where $\mathcal{F}_{ILR}$ indicates the loop filter network and $\theta$ encapsulates all of the learnable parameters in such network.

\subsection{Motion Decoupling: EEV-0.4}
When moving forward for further improvement, the EEV group considers the problem of essential motion representation in video coding. The key idea is to efficiently represent the motion between adjacent frames using an advanced model. In analogous to advanced motion vector prediction (AMVP) in conventional video coding~\cite{chien2021motion}, the decoupled motion model~\cite{lin2022dmvc} has been adopted by EEV-0.4 using the two-stage motion representation, namely MV prediction $m_t^i$ at frame $t$ and MV difference $m_t^c$. The former is formed by pair-wise hidden state $h_t^l$ and temporal context $c_t^l$ learned from the reference pictures using the convolutional long short-term memory (ConvLSTM) units, where $l$ specifies the number of stacked ConvLSTM units. Similarly, the MV difference is regarded as the combination of hidden state $h_t^\prime$ and temporal context $c_t^\prime$. $F$ refers to the transition of ConvLSTM units. Moreover, $H$ and $W$ are used to represent the height and width of the video.

\begin{figure}
    \centering
	\subfigure[ILR Network of EEV-0.4]{
	\includegraphics[width=0.17\textwidth]{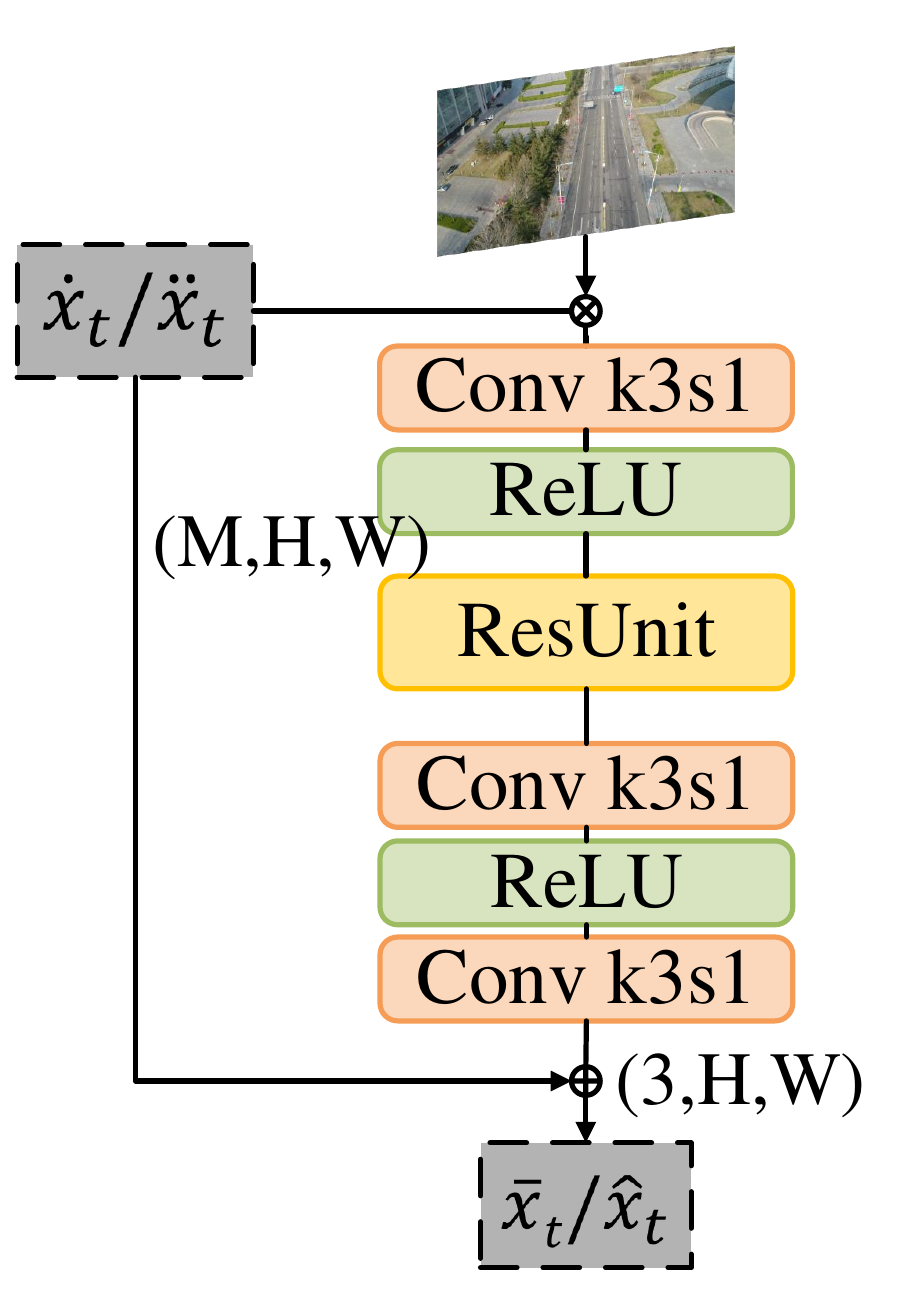}}
	\subfigure[Feature Extraction and Restoration]{
	\includegraphics[width=0.29\textwidth]{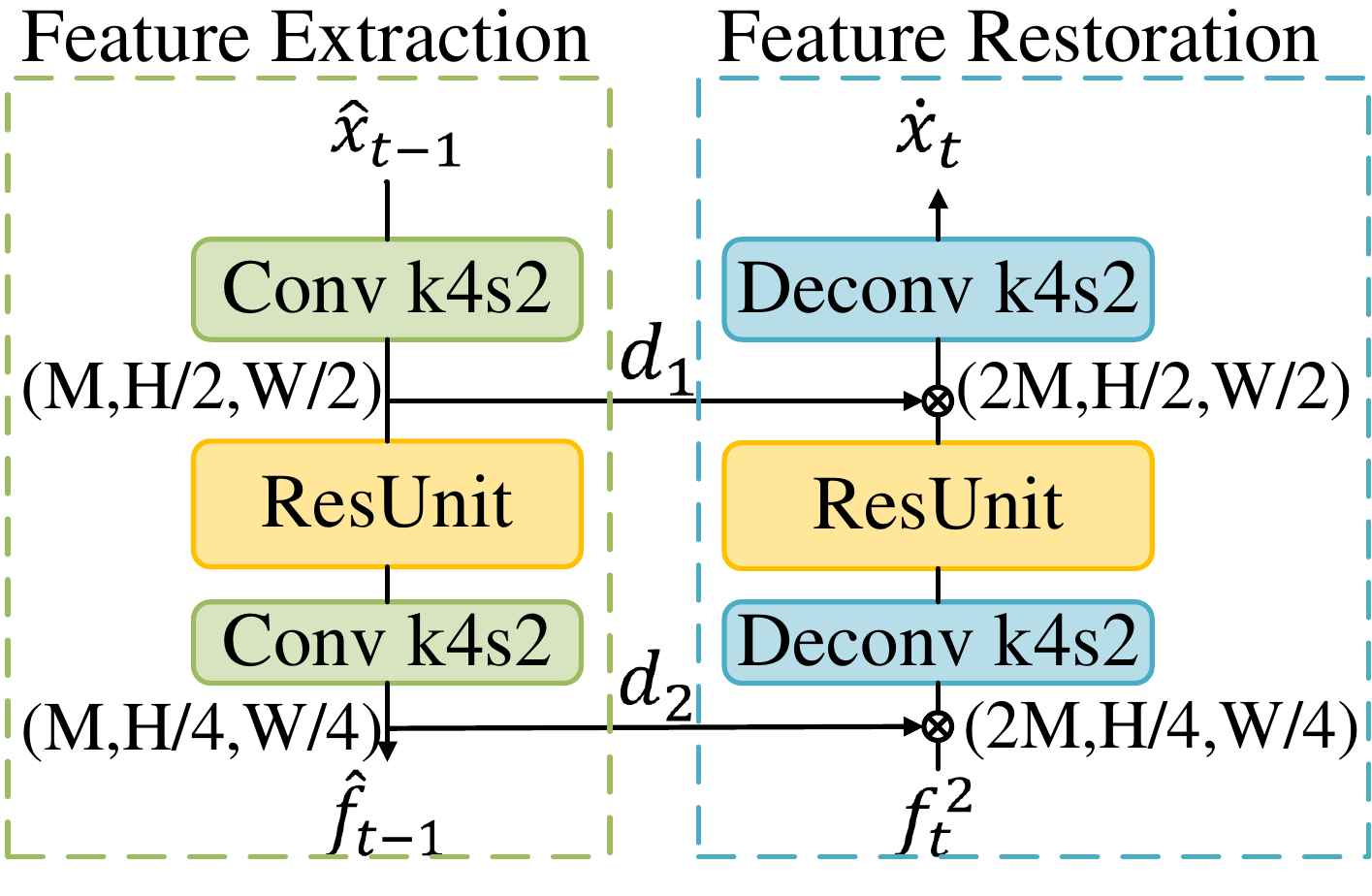}}
\caption{(a) Illustration of the spatiotemporal correlated residual network, which is employed for inter refinement and in-loop filter, and (b) the symmetric feature extraction and feature restoration module. 
}
\label{fig:framework}
\end{figure}

{\bf Feature Extraction and Feature Restoration.}
The overall description of the EEV-0.4 framework is illustrated in Fig.~\ref{fig:framework}. The forward feature transformation is responsible for representing $x_t$ as a compact deep feature representation. While the inverse feature transformation aims at restoring the $\hat{x}_t$ using such representation. The signaled bits are two folds, one for indicating the MV difference and the other for pixel-level residual. The entire workflow could be described as follows. The reconstructed pictures from the DPB are utilized as temporal contextual information to discover the MV prediction. All of the prediction operations are conducted in the feature space. As such, the reference picture with dimension of $(3, H, W)$ is embedded into the tensor with $(M, \frac{H}{4}, \frac{W}{4})$, where $M$ indicates the number of feature map. The downsampling process is then introduced to accommodate the state transition of large resolution frames during the inference of the ConvLSTM units. A symmetric architecture is designed for the feature extraction and feature restoration module, where the feature maps after each downsampling stage are attached to the feature restoration module. 

{\bf Progressive Inter Prediction.}
The motion decoupling is denoted as a progressive inter-prediction scheme, as illustrated in Fig.~\ref{fig:diagram-compare}(d). Relying on $m_t^i$, the temporal transition can be performed based on the stacked ConvLSTM units. The output $f_t^{1}$ of the last ConvLSTM unit corresponds to the coarse inter-prediction signals of time step $t$. To improve MV prediction, EEV-0.4 further provides the MV difference $m_t^c$. As shown in Fig.~\ref{fig:diagram-compare}(d), $m_t^c$ is unsupervisedly learned from pair-wise $\hat{x}_{t-1}$ and $x_t$, as the temporal transition is essentially occurred from $\hat{x}_{t-1}$ towards $x_t$. Note that approximated feature $f_t^1$ is merged with the decoded feature to infer $m_t^c$, which makes the network aware of the coarse temporal transition in the first place. With the guidance of MV difference $m_t^c$, the rough $f_t^{1}$ is enhanced to yield $f_t^2$, followed by a feature restoration net back to the pixel domain. 

{\bf Spatiotemporal Inter-pred Enhancement.}
As illustrated in Fig.~\ref{fig:framework}, the inter prediction $\bar{x}_t$ is further enhanced by a spatiotemporal refinement module as $\dot{x}_t$. Afterward, the pixel residue $r_t$ between $\bar{x}_t$ and $x_t$ is compressed and signaled. The variational-based image compression method with hyperprior entropy module~\cite{balle2018variational} is utilized for the residue coding. The reconstructed frame $\hat{x}_t$ can be viewed as the sum of restored residue $\hat{r}_t$ and $\bar{x}_t$. The ILR is the last step of the hybrid coding loop, producing the reconstruction $\hat{x}_t$. $\hat{x}_t$ is subsequently appended into the DPB for subsequent inter coding.

In summary, the technical features of the four EEV reference models are listed in Table~\ref{tab:coding-tool}. The MC and RC denote motion compensation and residual coding respectively. Interested audience could refer to~\cite{lin2022dmvc} for more details on the decomposed motion modeling of neural codecs.

\begin{table}[]
\centering
\caption{The coding tool list of EEV verification model}\scriptsize
\begin{tabular}{c|cccccc}
\hline
\textbf{Coding Tool} & \textbf{\begin{tabular}[c]{@{}c@{}}ME\\Net\end{tabular}} & \textbf{\begin{tabular}[c]{@{}c@{}}MC\\Net\end{tabular}} & \textbf{\begin{tabular}[c]{@{}c@{}}RC\\Module\end{tabular}} & \textbf{\begin{tabular}[c]{@{}c@{}}MCP\\Refine\end{tabular}} & \textbf{\begin{tabular}[c]{@{}c@{}}ILR\\Network\end{tabular}} & \textbf{\begin{tabular}[c]{@{}c@{}}Motion\\Decouple\end{tabular}} \\ \hline
EEV-0.1              & \Checkmark                                            & \Checkmark                                              & \Checkmark                                          & \XSolid                                                               & \XSolid                                                           & \XSolid                                                     \\
EEV-0.2              & \Checkmark                                            & \Checkmark                                              & \Checkmark                                          & \Checkmark                                                            & \XSolid                                                           & \XSolid                                                     \\
EEV-0.3              & \Checkmark                                            & \Checkmark                                              & \Checkmark                                          & \Checkmark                                                            & \Checkmark                                                        & \XSolid                                                     \\
EEV-0.4              & \Checkmark                                            & \Checkmark                                              & \Checkmark                                          & \Checkmark                                                            & \Checkmark                                                        & \Checkmark                                                 \\ \hline
\end{tabular}
\label{tab:coding-tool}
\end{table}

\section{Exploration Experiment of EEV}\label{EE}
This section presents the experiments of the verification models using aligned test conditions defined by MPAI-EEV and MPAI-EVC. The R-D performances of EEV-0.3 and EEV-0.4 models are mainly reported since the starting point of EEV lies in a publicly available model~\cite{yang2020opendvc}, which could be easily employed. Extensive comparisons for UAV video datasets are also carried out to illustrate the effectiveness of EEV models.

\subsection{Simulation Details}
{\bf Training strategy.} Extensive studies have been dedicated to the training processing of neural codecs. In EEV group, we simply follow the procedure defined in~\cite{yang2020opendvc,lu2019dvc} to train the EEV models. The vimeo-90K dataset~\cite{xue2019video} is chosen for the training and validation process. Regarding EEV-0.4, we follow the training protocol defined in~\cite{lin2022dmvc} to obtain the optimized parameters.

{\bf Testing dataset.} Testing datasets contain UAV video sequences~\cite{jia2023learning}, HEVC common test sequences, and UVG dataset. There are 14 video clips from UAV video sequences, which differ in terms of recording device type, location, environment, object, etc. The resolutions of them range from 640$\times$320 up to 2688$\times$1472 and each of them contains 100 frames. It’s important that all the videos have been center-cropped to be multiples of 64, making them compatible with convolution operations in learned codecs, especially for EEV-0.4. We additionally report the R-D efficiency using 15 HEVC common test videos for a better understanding EEV models.

{\bf Test configurations.} {The group of pictures (GOP) size and the intra-period number are set to 16 for all test sequences, which is slightly different from the common test condition (CTC) of HEVC and VVC. While we keep all other parameters to be the same with CTCs except for the GOP size and intra-period number. The hyper-parameter $\lambda$ balances the trade-off between coding bits and distortion for all learned video codecs. As for PSNR metric, the $\lambda$ is set to be \{2048, 1024, 512, 256\}, while for MS-SSIM metric, $\lambda$ values are \{64, 32, 16, 8\}.}

Regarding EEV-0.4, the first frame (I-picture) of each GOP is coded by the pre-trained E2E image compression model~\cite{cheng2020learned} implemented by CompressAI~\cite{begaint2020compressai}. The quality levels are set to \{3, 4, 5, 6\}, respectively, for different bit-rate coding scenarios. Note that the mode and metric parameters are set to be either mse or msssim. The selected pre-trained model should be consistent with pre-defined $\lambda$s. The detailed command is shown as follows.

\begin{itemize}
\item \textit{python eval.py --eval\_lambda $\lambda$ --metric mse --intra\_model cheng2020\_anchor --test\_class seq\_Class --gop\_size 16 --pretrain ./checkpoints/dmvc\_psnr\_$\lambda$.model}
\item \textit{python eval.py --eval\_lambda $\lambda$ --metric ms-ssim --intra\_model cheng2020\_anchor --test\_class seq\_Class --gop\_size 16 --pretrain ./checkpoints/dmvc\_msssim\_$\lambda$.model}
\end{itemize}

Regarding EEV-0.3, the first intra frame is coded by BPG for PSNR metric and the pre-trained model with context-adaptive entropy model released by~\cite{lee2018context} for MS-SSIM metric. The quality levels for PSNR are set to 22, 27, 32, and 37, while for MS-SSIM, they are set to 2, 3, 5, and 7. The remaining P frames are compressed sequentially by such model, with the coding mode and metric parameter set as follows.

\begin{itemize}
\item \textit{python test\_eev.py -path sequence\_name -mode PSNR -IntraPeriod 16 -metric PSNR -l $\lambda$}
\end{itemize}

To evaluate the EEV-0.3 and EEV-0.4 reference models, HEVC, EEV-0.1, and VVC are also used for comprehensive comparisons. For HEVC standard, we employ the HEVC screen content coding (HEVC-SCC) extension~\cite{xu2015overview} reference software (HM-16.20-SCM-8.8). The InputColourSpaceConvert parameter is set as RGBtoGBR for encoding, and the OutputInternalColourSpace is set as GBRtoRGB for decoding to ensure the consistency of the reconstruction results and decoded results. The internal bit-depth is set to be 8, and the quantization parameters (QPs) are set to 30, 34, 38, and 42 to achieve four different bit-rate points. The specific command is provided below.

\begin{itemize}
\item \textit{TAppEncoder -c encoder\_lowdelay\_P\_main.cfg -InputBitDepth 8
 -InputChromaFormat 444 -Level 6.2 -wdt seq\_wid -hgt seq\_hgt -f 100 -fr fps -q QP -IntraPeriod 16 -InputColourSpaceConvert RGBtoGBR -SNRInternalColourSpace 1\\ -OutputColourSpaceConvert GBRtoRGB}
\end{itemize}

To illustrate the performance of H.266/VVC standard, we adopt the reference software VTM-15.2 as a solid benchmark to evaluate the performance of these sequences. The specific parameter-setting command is provided below.

\begin{itemize}
\item \textit{EncoderAppStatic -c ./cfg/encoder\_lowdelay\_vtm.cfg -i input.yuv –InputBitDepth=8 –OutputBitDepth=8 –DecodingRefreshType=2 -f 100 -q QP -fr fr -wdt W -hgt H –IntraPeriod=16 -o output.yuv}
\end{itemize}
where $W$ and $H$ stand for the image height and width, $fr$ denotes frame rate. The variable-rate coding is also controlled by the $\lambda$s in EEV-0.1. Two different sets of values for $\lambda$ are \{2048,1024,512,256\} for PSNR and \{64,32,16,8\} for MS-SSIM~\cite{wang2003multiscale}, respectively. The coding mode and evaluation metric parameter settings remain the same with EEV-0.3. The specific command is provided below.

\begin{itemize}
\item \textit{python test\_opendvc.py -path sequence\_name -mode PSNR -IntraPeriod 16 -metric PSNR -l $\lambda$}
\end{itemize}

\begin{table*}[]
\centering
\footnotesize
\caption{The BD-rate performance of different codecs (EEV-0.4, EEV-0.3, VTM-15.2 and HM-16.20-SCM-8.8) on drone video compression. The distortion metric is RGB-MS-SSIM.}
\begin{tabular}{cc|c|c|c|c}
\hline
\multicolumn{1}{c|}{\textbf{Category}}                                                               & \textbf{\begin{tabular}[c]{@{}c@{}}Sequence\\ Name\end{tabular}} & \textbf{\begin{tabular}[c]{@{}c@{}}BD-Rate Reduction\\ EEV-0.4 vs HEVC\end{tabular}} & \textbf{\begin{tabular}[c]{@{}c@{}}BD-Rate Reduction\\ EEV-0.3 vs HEVC\end{tabular}} & \textbf{\begin{tabular}[c]{@{}c@{}}BD-Rate Reduction\\ VVC vs HEVC\end{tabular}} & \textbf{\begin{tabular}[c]{@{}c@{}}BD-Rate Reduction\\ EEV-0.1 vs HEVC\end{tabular}} \\ \hline
\multicolumn{1}{c|}{\multirow{5}{*}{\begin{tabular}[c]{@{}c@{}}Class A\\ VisDrone-SOT~\cite{zhu2021detection}\end{tabular}}} & BasketballGround & -49.23\% & 43.93\% & -44.40\% & 4.73\%\\ 
\multicolumn{1}{c|}{} & GrassLand & -70.38\% & 2.83\%  & -61.42\% & -54.31\%\\ 
\multicolumn{1}{c|}{} & Intersection & -71.72\% & -22.99\%  & -57.51\% & -49.96\% \\ 
\multicolumn{1}{c|}{} & NightMall & -61.60\% & -12.62\%  & -46.53\% & -33.28\%\\ 
\multicolumn{1}{c|}{} & SoccerGround & -66.36\% & -15.83\%  & -59.58\% & -55.47\%\\ \hline
\multicolumn{1}{c|}{\multirow{3}{*}{\begin{tabular}[c]{@{}c@{}}Class B\\ VisDrone-MOT~\cite{zhu2021detection}\end{tabular}}} & Circle & -69.13\% & -19.28\%  & -57.98\% & -52.36\%\\ 
\multicolumn{1}{c|}{} & CrossBridge & -45.78\% & 0.29\%  & -44.59\% & -18.44\%\\ 
\multicolumn{1}{c|}{} & Highway & -59.94\% & -29.13\%  & -55.25\% & -33.32\%\\ \hline
\multicolumn{1}{c|}{\multirow{3}{*}{\begin{tabular}[c]{@{}c@{}}Class C\\ Corridor~\cite{kouris2019informed}\end{tabular}}} & Classroom & -61.67\% & -21.13\%  & -85.60\% & -32.49\%\\ 
\multicolumn{1}{c|}{} & Elevator & -80.74\% & -54.63\%  & -79.79\% & -57.37\%\\ 
\multicolumn{1}{c|}{} & Hall & -73.56\% & -32.06\%  & -77.71\% & -51.84\%\\ \hline
\multicolumn{1}{c|}{\multirow{3}{*}{\begin{tabular}[c]{@{}c@{}}Class D\\ UAVDT\_S~\cite{du2018unmanned}\end{tabular}}} & Campus & -64.56\% & -28.87\%  & -51.61\% & -41.51\% \\ 
\multicolumn{1}{c|}{} & RoadByTheSea & -64.09\% & -27.77\%  & -54.15\% & -39.97\%\\ 
\multicolumn{1}{c|}{} & Theater & -51.73\% & 6.59\% & -59.72\% & -20.14\%\\ \hline
\multicolumn{2}{c|}{\textbf{Class A}} & \textbf{-63.86\%} & \textbf{-0.94\%} & \textbf{-53.89\%} & \textbf{-37.66\%} \\ 
\multicolumn{2}{c|}{\textbf{Class B}} & \textbf{-58.29\%} & \textbf{-16.04\%} & \textbf{-52.60\%} & \textbf{-34.70\%}  \\ 
\multicolumn{2}{c|}{\textbf{Class C}} & \textbf{-71.99\%} & \textbf{-35.94\%} & \textbf{-81.03\%} & \textbf{-47.23\%} \\ 
\multicolumn{2}{c|}{\textbf{Class D}} & \textbf{-60.13\%} & \textbf{-16.68\%} & \textbf{-55.16\%} & \textbf{-33.87\%} \\ \hline
\multicolumn{2}{c|}{\textbf{Average}} & \textbf{-63.61\%} & \textbf{-15.05\%} & \textbf{-59.70\%} & \textbf{-38.26\%} \\ \hline
\end{tabular}
\label{tab:rdperformance-msssim}
\end{table*}

\begin{figure}
    \centering
	\subfigure[Intersection]{
	\includegraphics[width=0.23\textwidth]{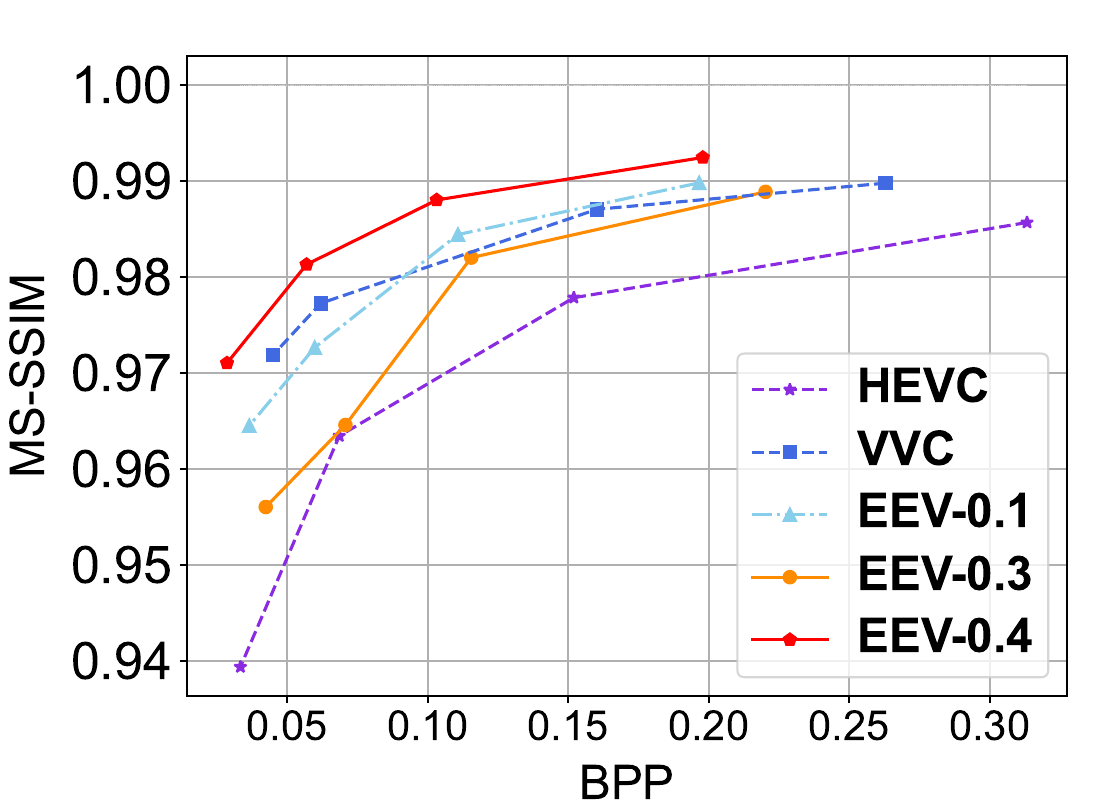}}
	\subfigure[Highway]{
	\includegraphics[width=0.23\textwidth]{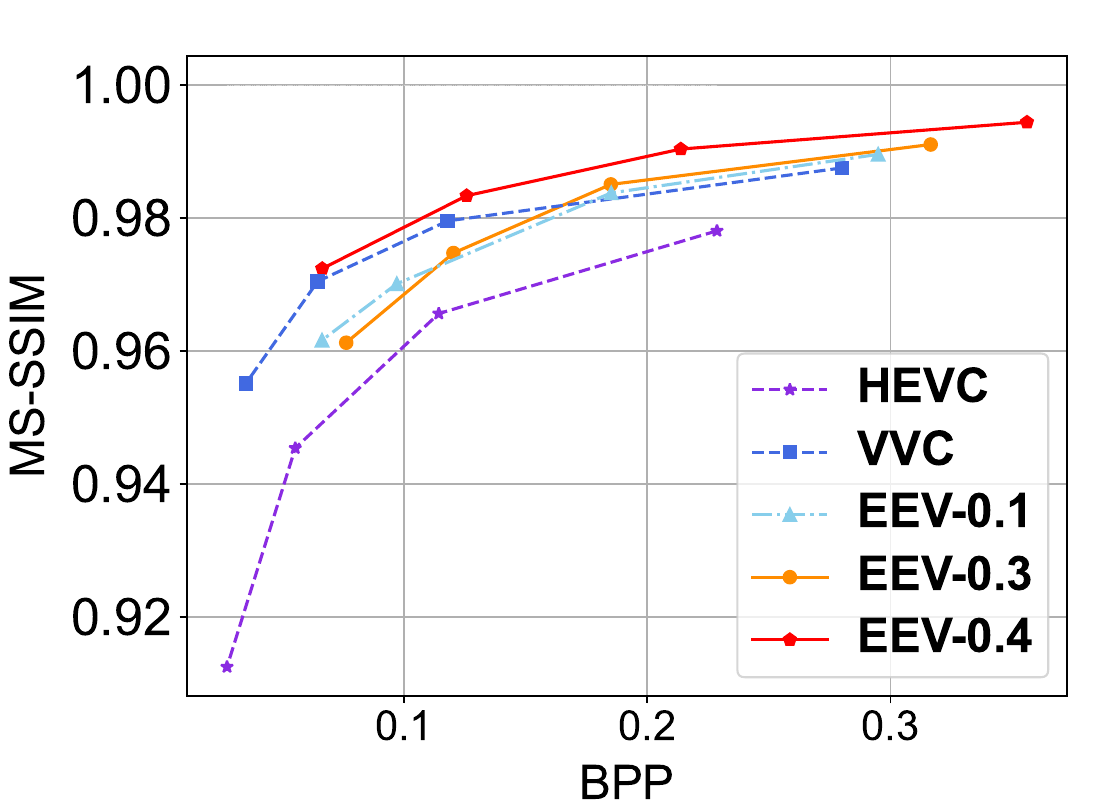}}
	\subfigure[Hall]{
	\includegraphics[width=0.23\textwidth]{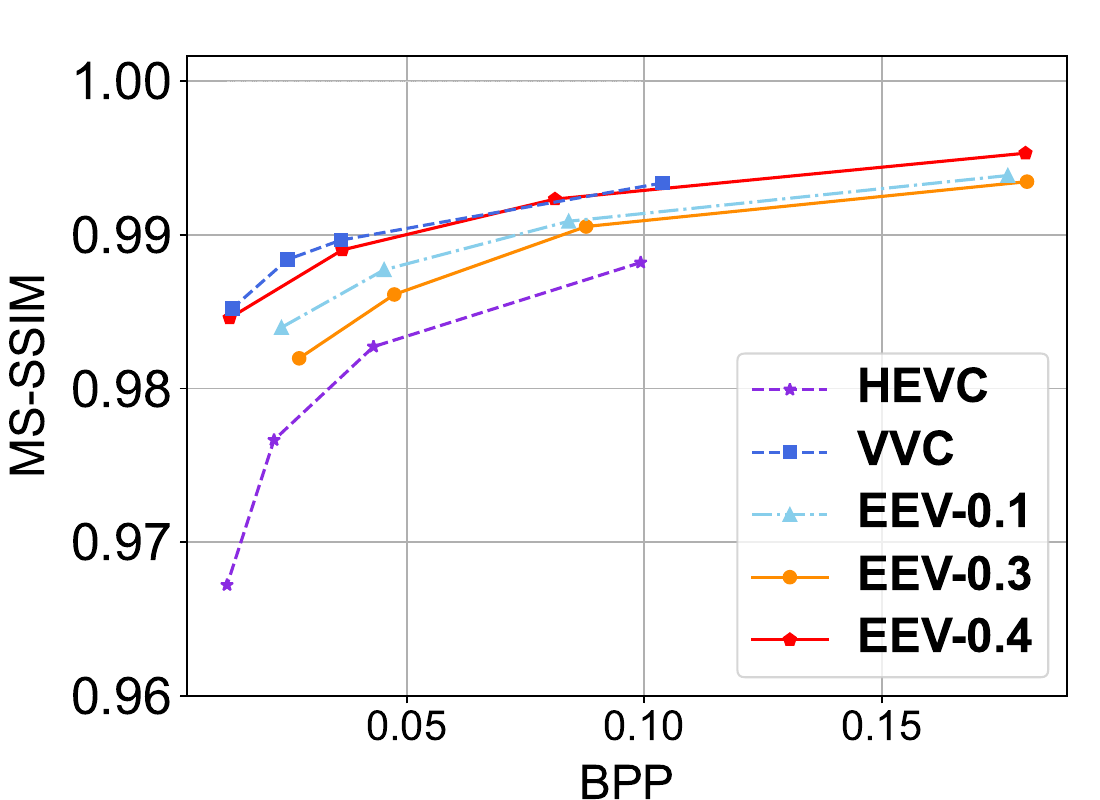}}
	\subfigure[Campus]{
	\includegraphics[width=0.23\textwidth]{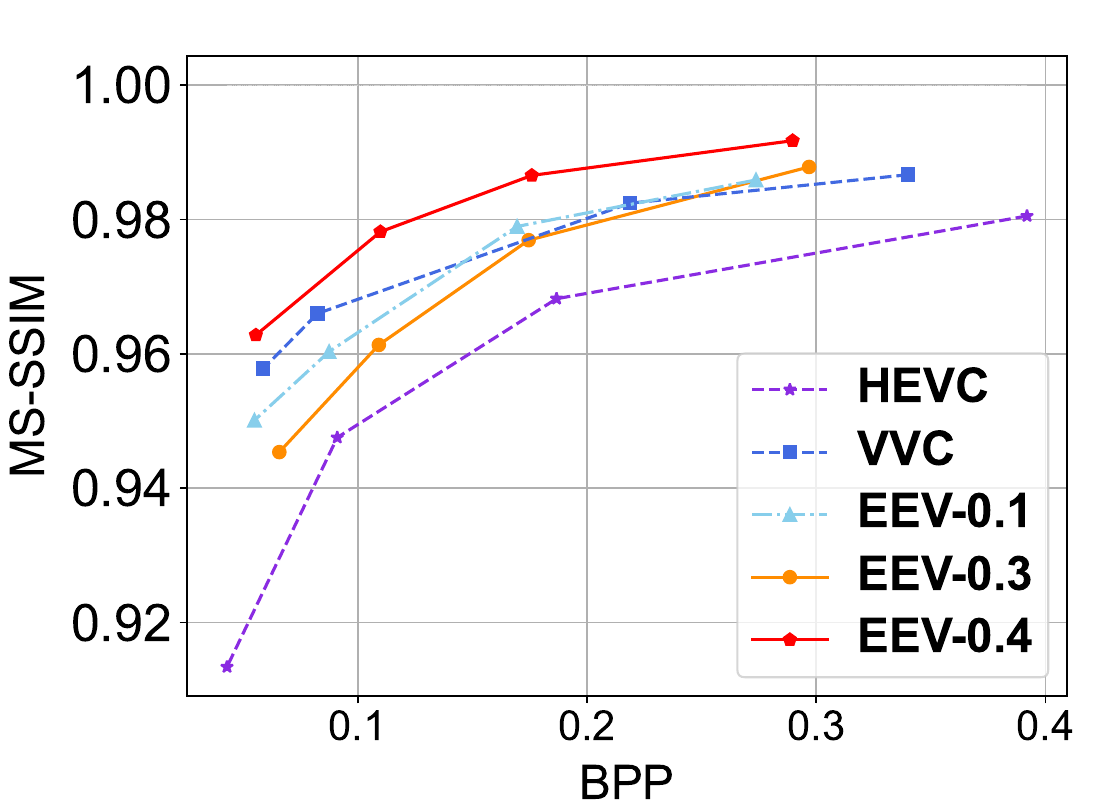}}
\caption{The R-D curves of the four UAV test sequences using MS-SSIM metric. The performances of HEVC, VVC, EEV-0.1, EEV-0.3, and EEV-0.4 are depicted.}
\label{fig:rdcurves_msssim}
\end{figure}

{\bf MS-SSIM Metric.} Taking BPP and reconstructed MS-SSIM values as the horizontal and vertical axes, respectively, the R-D curves of different coding methods are plotted in Fig.~\ref{fig:rdcurves_msssim}. The detailed experimental results are provided in Table.~\ref{tab:rdperformance-msssim}. The BD-rate performances and R-D curves show that EEV-0.4 significantly outperforms HEVC for all sequences, demonstrating our method's stronger generalization ability and adaptability to complex scenarios. Moreover, EEV-0.3 is inferior to HEVC on some sequences but performs better when averaging all sequences, with 15.05\% BD-rate saving compared to HEVC reference software.

\begin{figure*}
    \centering
	\subfigure[EEV-0.4 (0.0703 / 0.9569 / 26.9356)]{
	\includegraphics[width=0.32\textwidth]{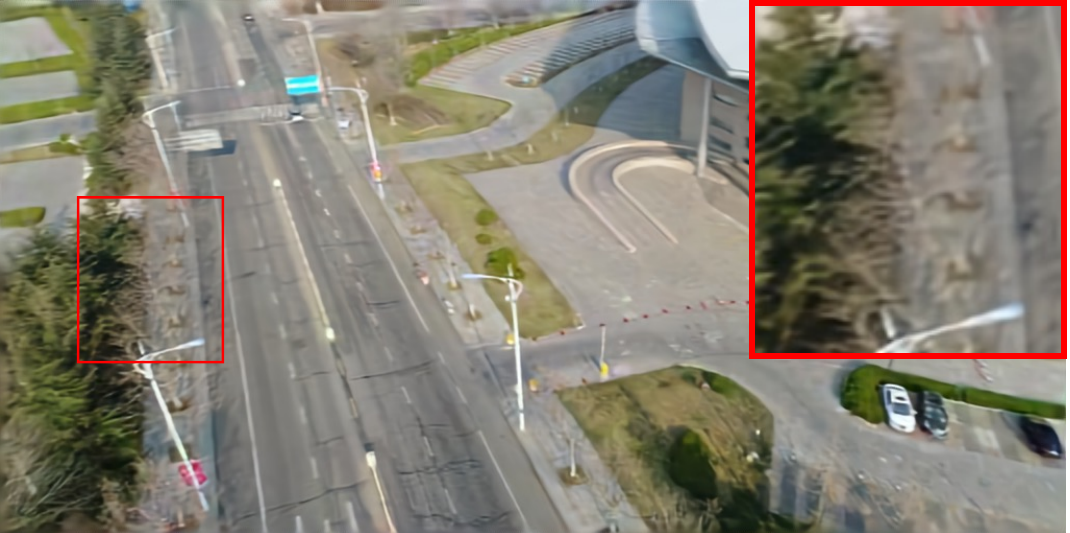}}
	\subfigure[VVC (0.0753 / 0.9657 / 30.2692)]{
	\includegraphics[width=0.32\textwidth]{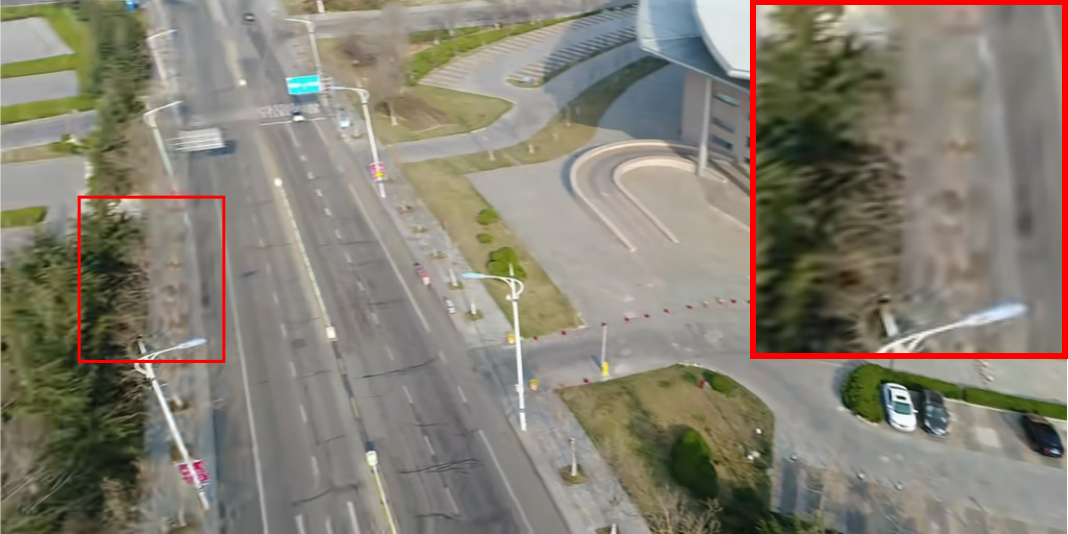}}
	\subfigure[Uncompressed]{
	\includegraphics[width=0.32\textwidth]{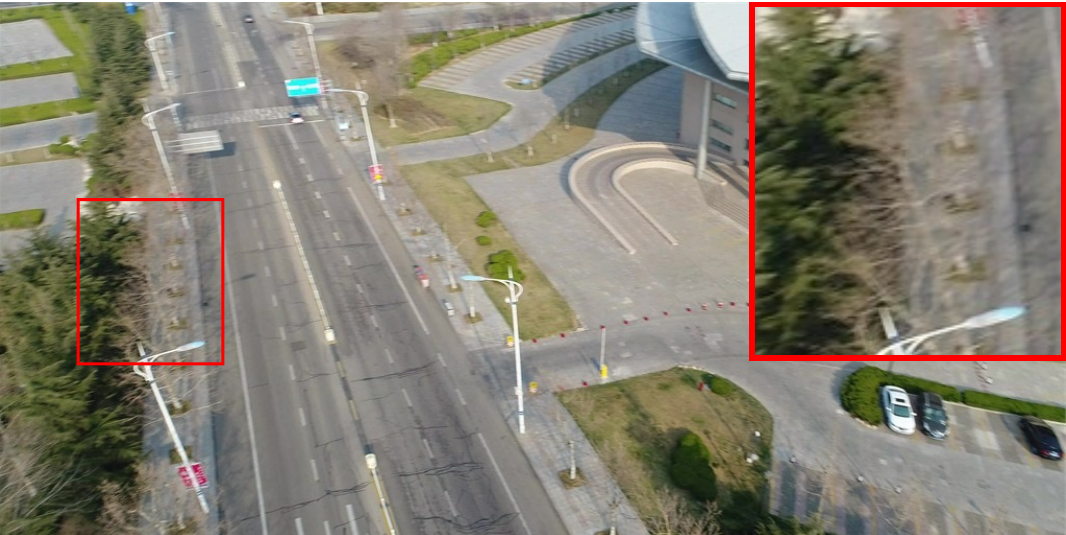}}
	\subfigure[EEV-0.4 (0.0326 / 0.9772 / 30.2538)]{
	\includegraphics[width=0.32\textwidth]{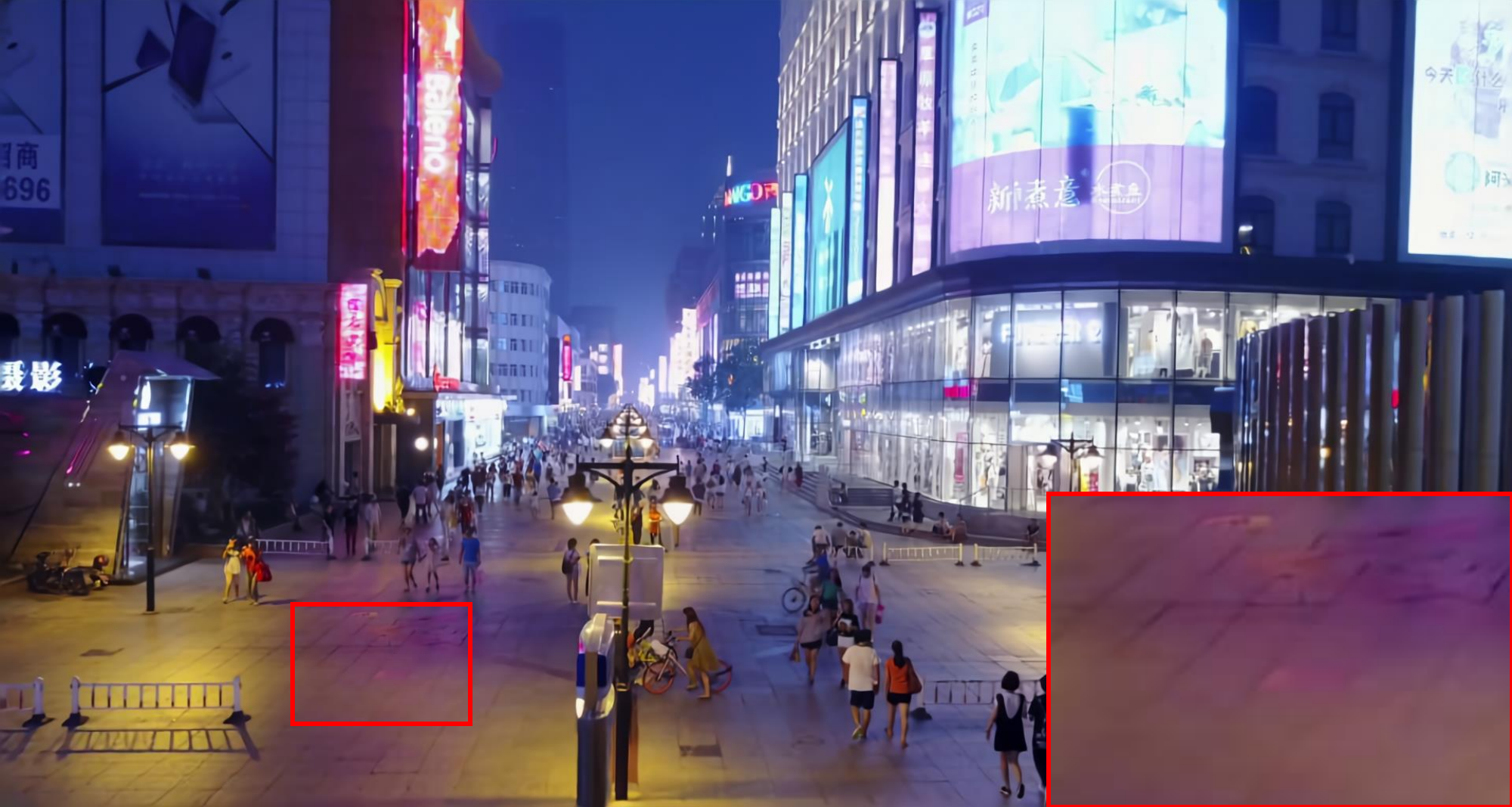}}
	\subfigure[VVC (0.0388 / 0.9756 / 32.6756)]{
	\includegraphics[width=0.32\textwidth]{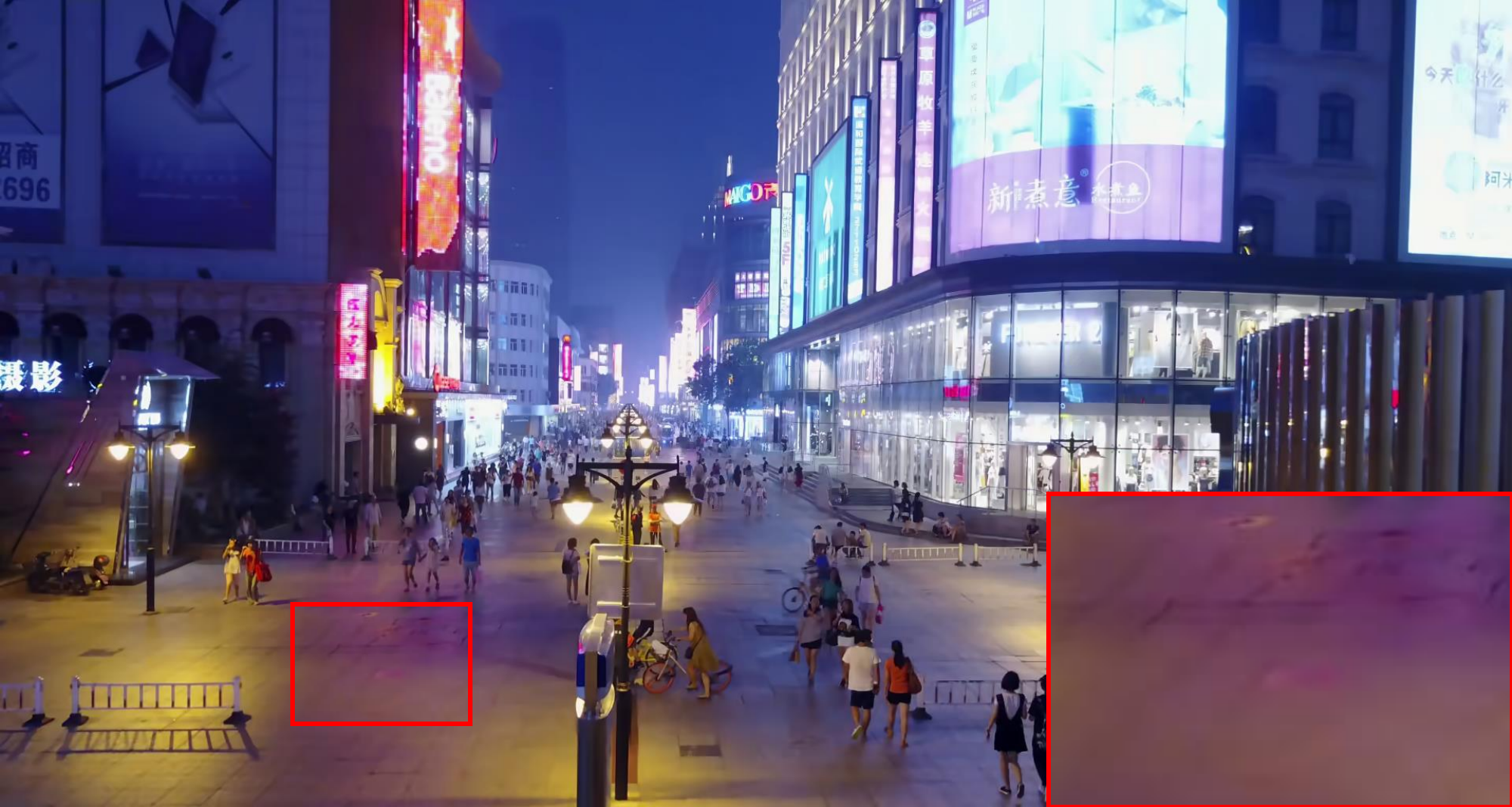}}
	\subfigure[Uncompressed]{
	\includegraphics[width=0.32\textwidth]{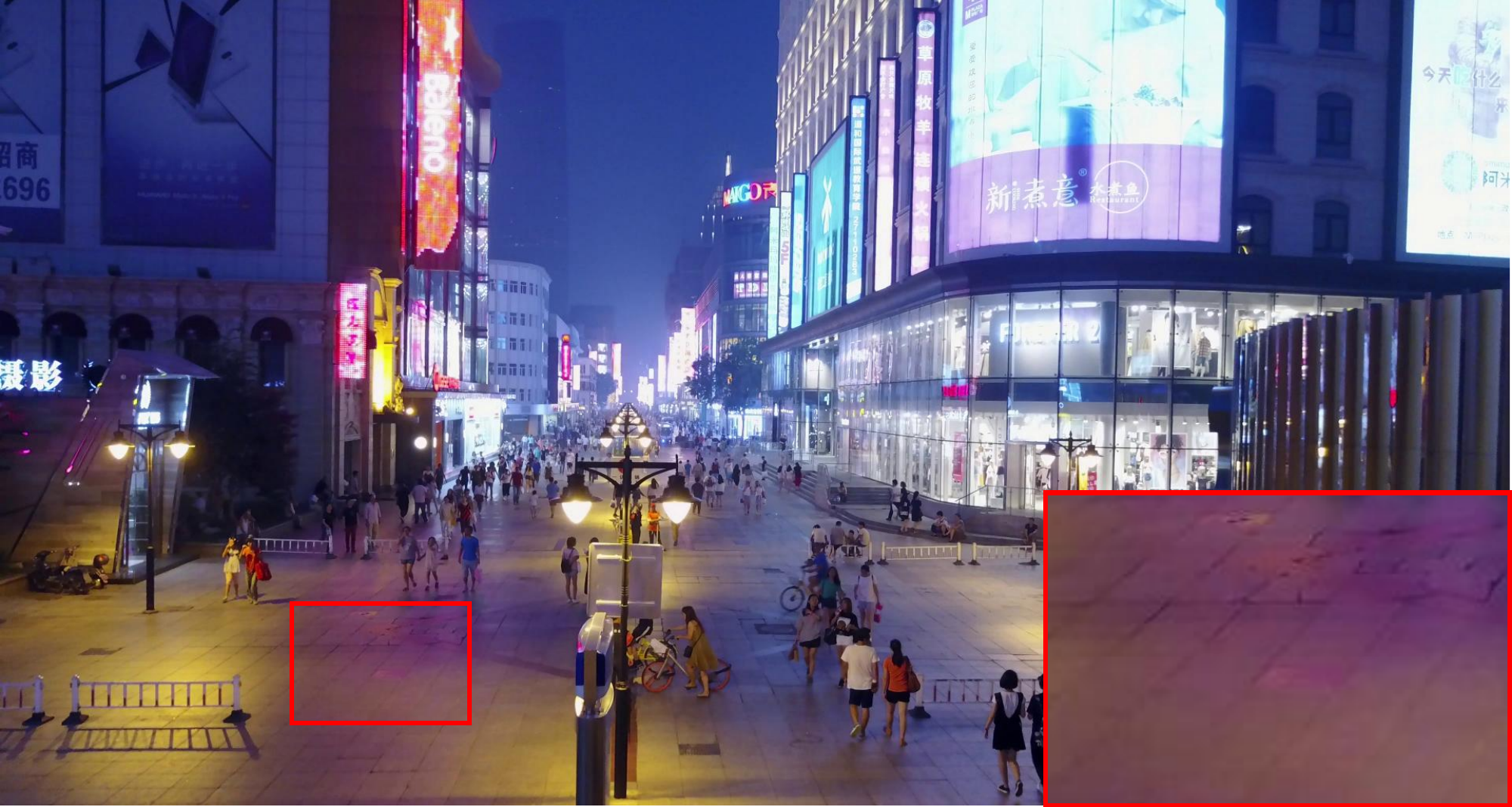}}
	\subfigure[EEV-0.4 (0.0241 / 0.9738 / 31.1073)]{
	\includegraphics[width=0.32\textwidth]{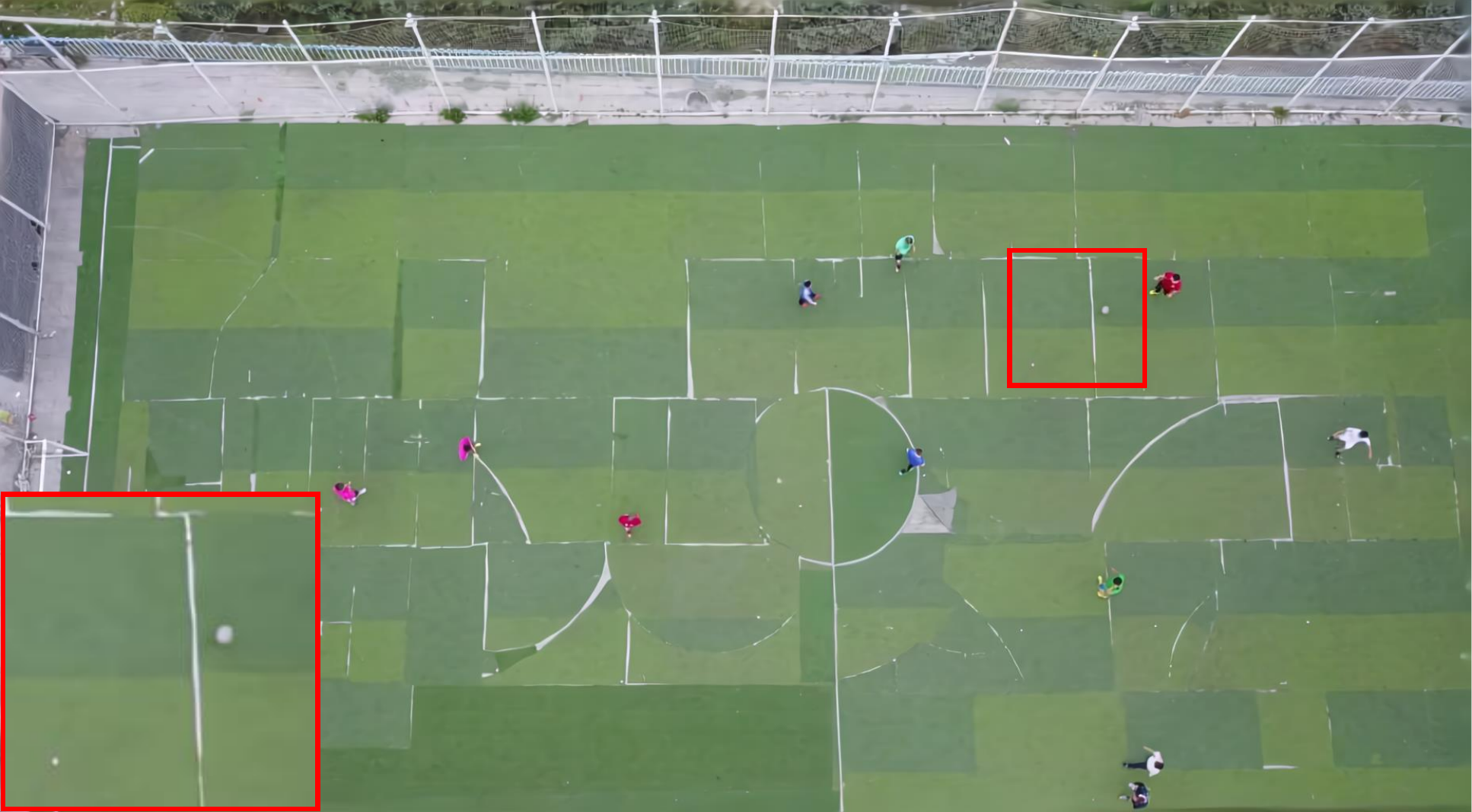}}
	\subfigure[VVC (0.0235 / 0.9691 / 35.0742)]{
	\includegraphics[width=0.32\textwidth]{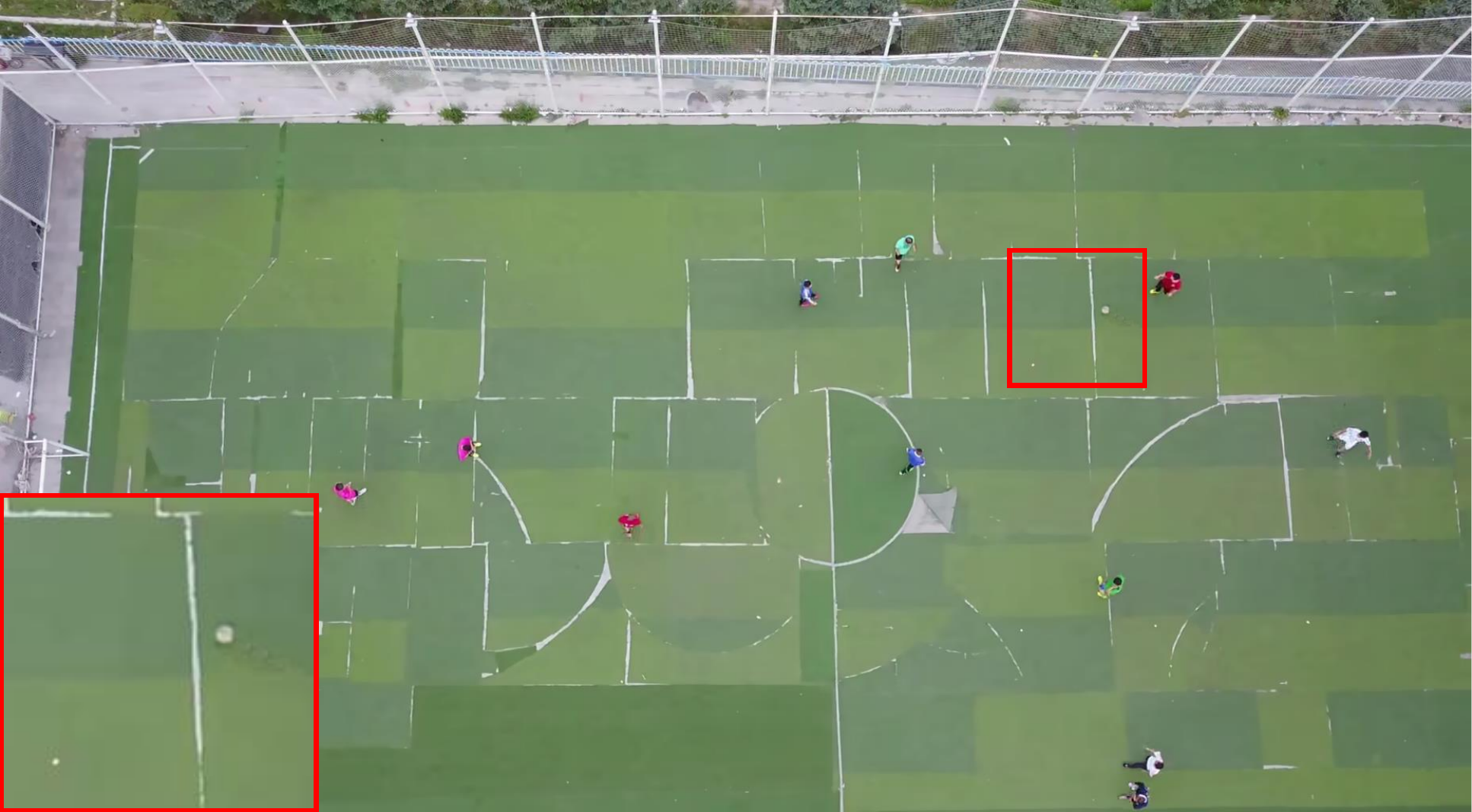}}
	\subfigure[Uncompressed]{
	\includegraphics[width=0.32\textwidth]{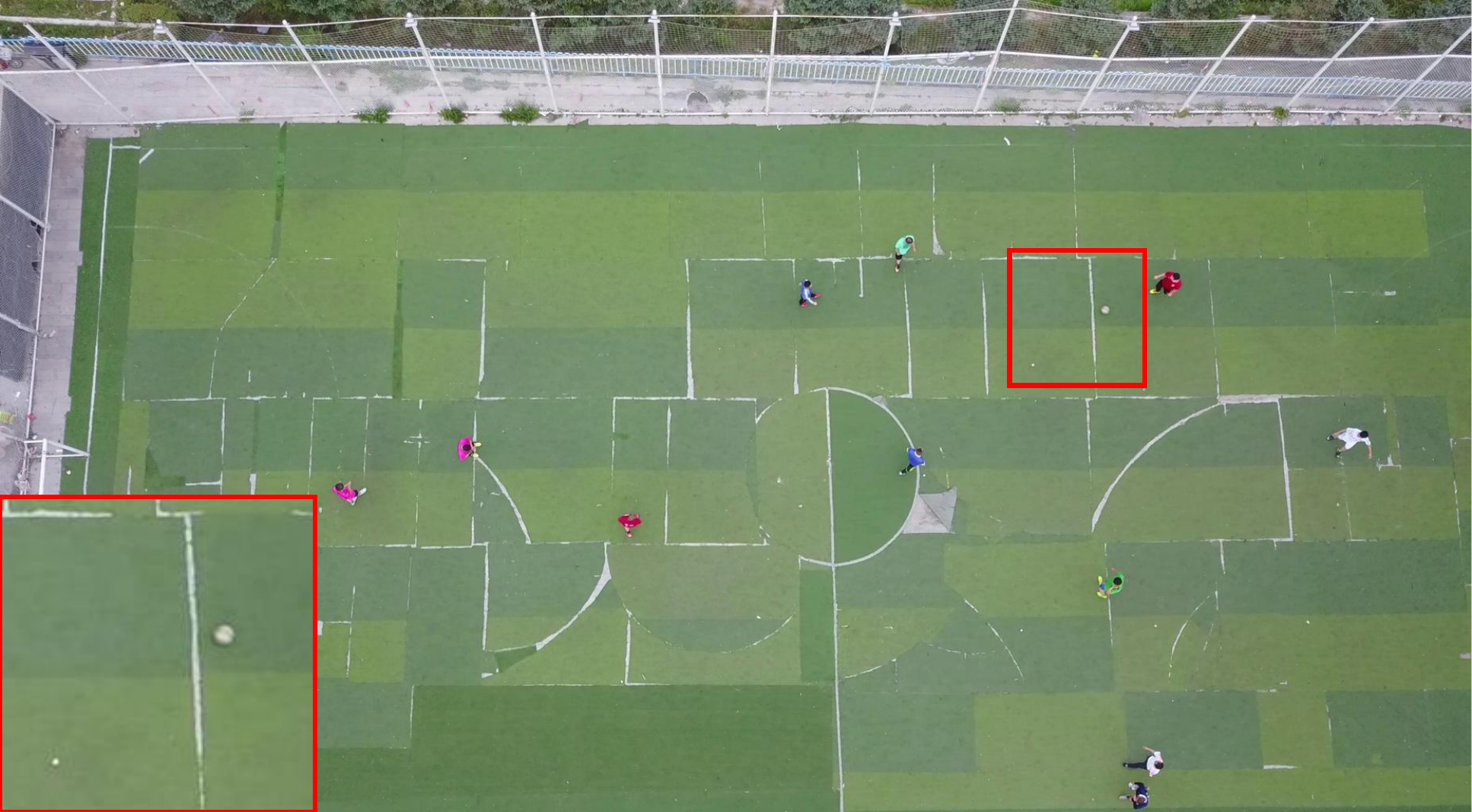}}
\caption{The subjective quality comparison of {\bf BasketballGround}, {\bf NightMall} and {\bf SoccerGround} sequence for the EEV-0.4 model, H.266/VVC, and their corresponding pristine frames. The corresponding \textit{BPP}, \textit{MS-SSIM}, and \textit{PSNR} values are provided below each subfigure.}
\label{fig:visual-comparison}
\end{figure*}

Regarding performance comparison with VVC, as shown in Fig.~\ref{fig:rdcurves_msssim}, EEV-0.3 dominates the performances only at high bitrates, while EEV-0.4 always has the highest efficiency in most bitrate range. To better demonstrate the relative quality, we selected restored pictures shown in Fig.~\ref{fig:visual-comparison}. Specifically, we choose the case where the QP for VVC is 40. The corresponding $\lambda$ is selected according to BPP for EEV-0.4. EEV-0.4 achieves higher visual quality with more complex textures. 

\begin{table*}[]
\centering
\footnotesize
\caption{The BD-rate performance of different codecs (EEV-0.4, EEV-0.3, VTM-15.2 and HM-16.20-SCM-8.8) on drone video compression. The distortion metric is RGB-PSNR.}
\begin{tabular}{cc|c|c|c|c}
\hline
\multicolumn{1}{c|}{\textbf{Category}}                                                               & \textbf{\begin{tabular}[c]{@{}c@{}}Sequence\\ Name\end{tabular}} & \textbf{\begin{tabular}[c]{@{}c@{}}BD-Rate Reduction\\ EEV-0.4 vs HEVC\end{tabular}} & \textbf{\begin{tabular}[c]{@{}c@{}}BD-Rate Reduction\\ EEV-0.3 vs HEVC\end{tabular}} & \textbf{\begin{tabular}[c]{@{}c@{}}BD-Rate Reduction\\ VVC vs HEVC\end{tabular}} & \textbf{\begin{tabular}[c]{@{}c@{}}BD-Rate Reduction\\ EEV-0.1 vs HEVC\end{tabular}} \\ \hline
\multicolumn{1}{c|}{\multirow{5}{*}{\begin{tabular}[c]{@{}c@{}}Class A\\ VisDrone-SOT~\cite{zhu2021detection}\end{tabular}}} & BasketballGround & -27.11\% & 9.04\% & -44.22\% & 44.00\% \\ 
\multicolumn{1}{c|}{} & GrassLand & -51.61\% & -38.90\%  & -58.63\% & -25.02\% \\ 
\multicolumn{1}{c|}{} & Intersection & -54.54\% & -29.20\%  & -56.57\% & -12.41\% \\ 
\multicolumn{1}{c|}{} & NightMall & -42.19\% & -7.43\%  &-50.21\%  & 19.39\% \\ 
\multicolumn{1}{c|}{} & SoccerGround & -44.68\% & -11.37\%  & -61.46\% & 16.39\% \\ \hline
\multicolumn{1}{c|}{\multirow{3}{*}{\begin{tabular}[c]{@{}c@{}}Class B\\ VisDrone-MOT~\cite{zhu2021detection}\end{tabular}}} & Circle & -36.34\% & -26.72\%  & -55.64\%  & -7.53\% \\ 
\multicolumn{1}{c|}{} & CrossBridge & -6.37\% & 28.70\%  & -61.58\% & 71.23\%  \\ 
\multicolumn{1}{c|}{} & Highway & -40.09\% & -14.23\%  & -56.91\% & 8.26\%  \\ \hline
\multicolumn{1}{c|}{\multirow{3}{*}{\begin{tabular}[c]{@{}c@{}}Class C\\ Corridor~\cite{kouris2019informed}\end{tabular}}} & Classroom & -39.01\% & 77.69\%  & -45.06\% & 63.97\%  \\ 
\multicolumn{1}{c|}{} & Elevator & -41.27\% & 18.64\%  & -40.16\% & 28.92\%  \\ 
\multicolumn{1}{c|}{} & Hall & -48.13\% & -4.66\%  & -42.95\% & 7.47\%  \\ \hline
\multicolumn{1}{c|}{\multirow{3}{*}{\begin{tabular}[c]{@{}c@{}}Class D\\ UAVDT\_S~\cite{du2018unmanned}\end{tabular}}} & Campus & -50.31\% & -26.19\%  & -52.55\% & 2.48\%  \\ 
\multicolumn{1}{c|}{} & RoadByTheSea & -44.76\% & -24.88\%  & -54.95\% & -3.72\%  \\ 
\multicolumn{1}{c|}{} & Theater & -29.69\% & 3.31\% & -57.50\% & 33.57\%  \\ \hline
\multicolumn{2}{c|}{\textbf{Class A}} & \textbf{-44.02\%} & \textbf{-15.57\%} & \textbf{-54.22\%} & \textbf{8.47\%} \\
\multicolumn{2}{c|}{\textbf{Class B}} & \textbf{-27.60\%} & \textbf{-4.08\%} & \textbf{-58.04\%}  & \textbf{23.99\%} \\ 
\multicolumn{2}{c|}{\textbf{Class C}} & \textbf{-42.80\%} & \textbf{30.56\%} & \textbf{-42.73\%} & \textbf{33.45\%} \\ 
\multicolumn{2}{c|}{\textbf{Class D}} & \textbf{-41.58\%} & \textbf{-15.91\%} & \textbf{-55.00\%} & \textbf{10.78\%} \\ \hline
\multicolumn{2}{c|}{\textbf{Average}} & \textbf{-39.72\%} & \textbf{-3.30\%} & \textbf{-52.74\%} & \textbf{17.64\%} \\ \hline
\end{tabular}
\label{tab:rdperformance-psnr}
\end{table*}

\begin{figure*}
    \centering
	\subfigure[Intersection]{
	\includegraphics[width=0.23\textwidth]{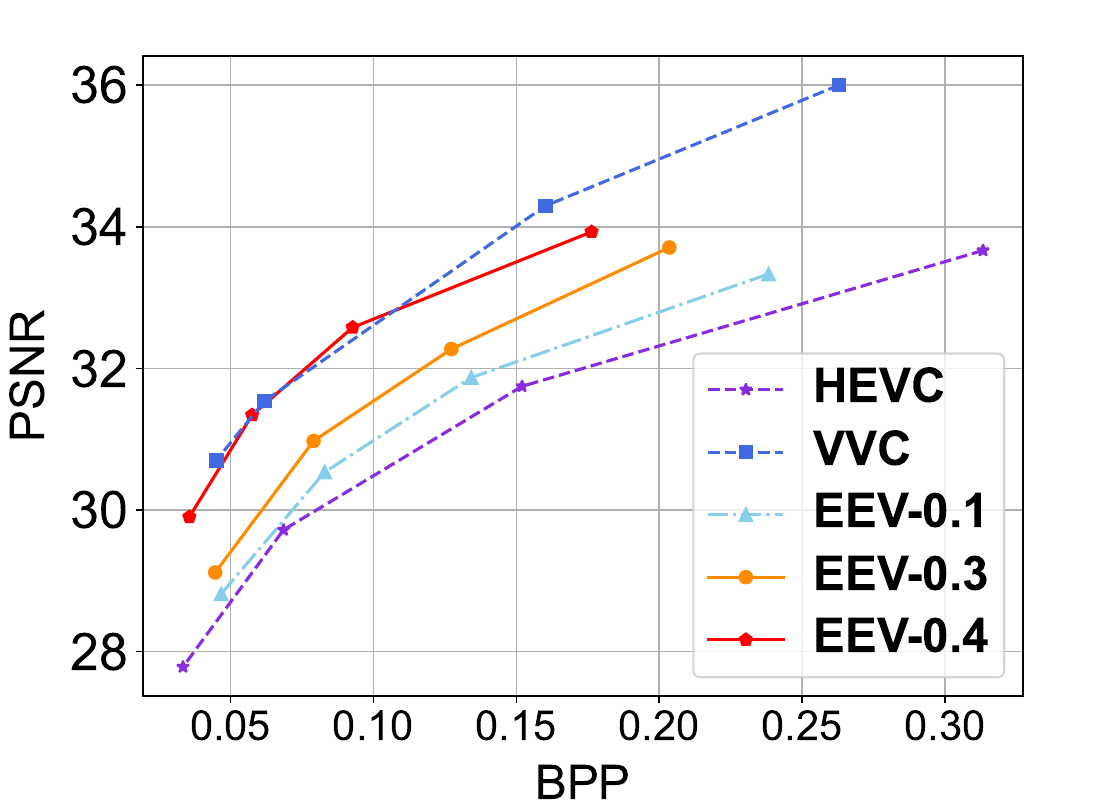}}
	\subfigure[Highway]{
	\includegraphics[width=0.23\textwidth]{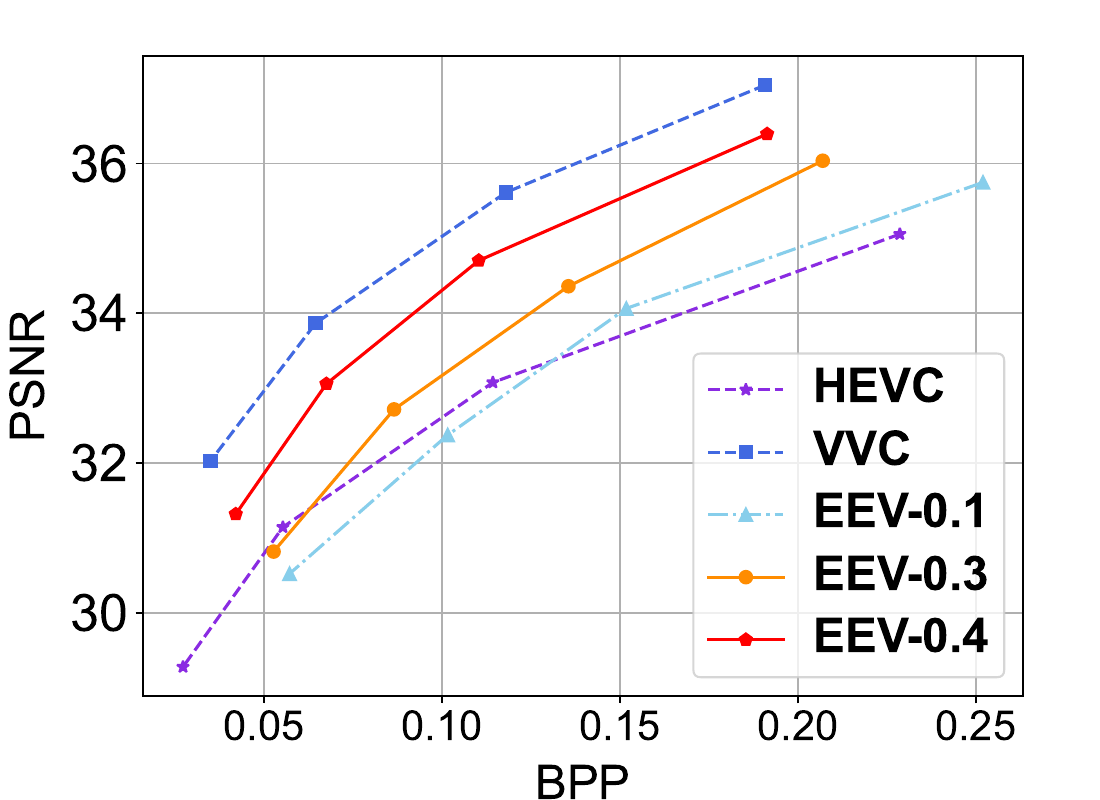}}
	\subfigure[Hall]{
	\includegraphics[width=0.23\textwidth]{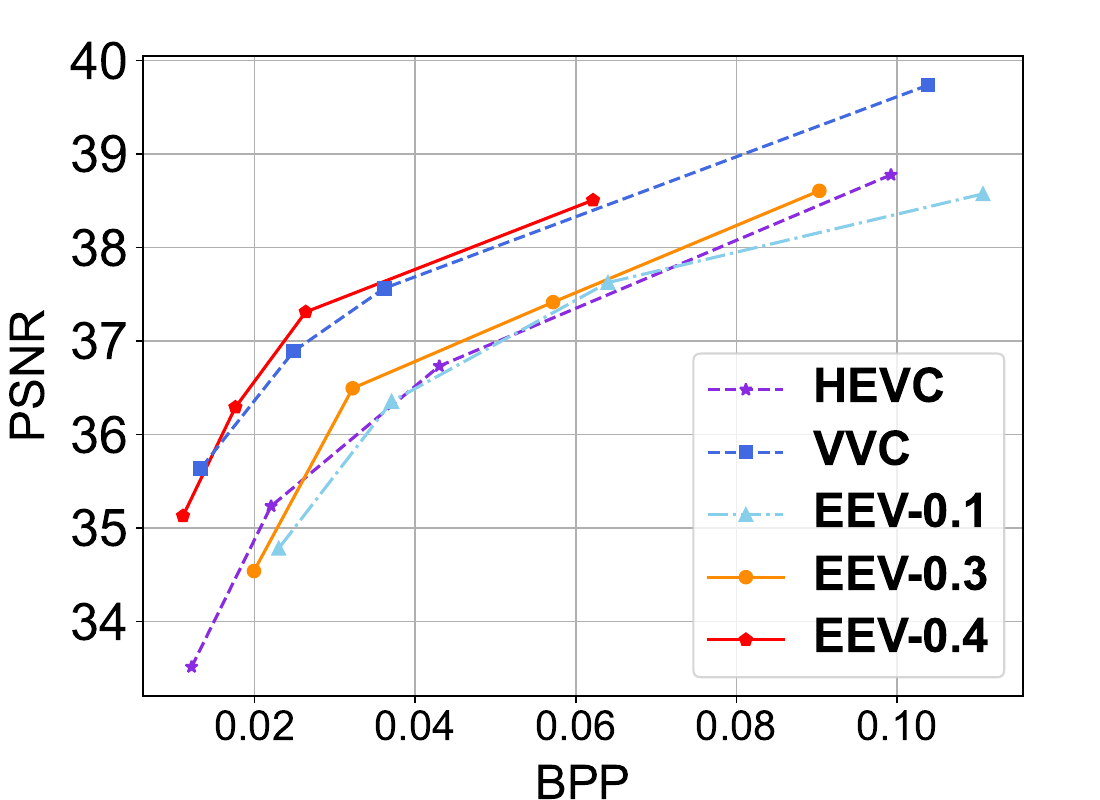}}
	\subfigure[Campus]{
	\includegraphics[width=0.23\textwidth]{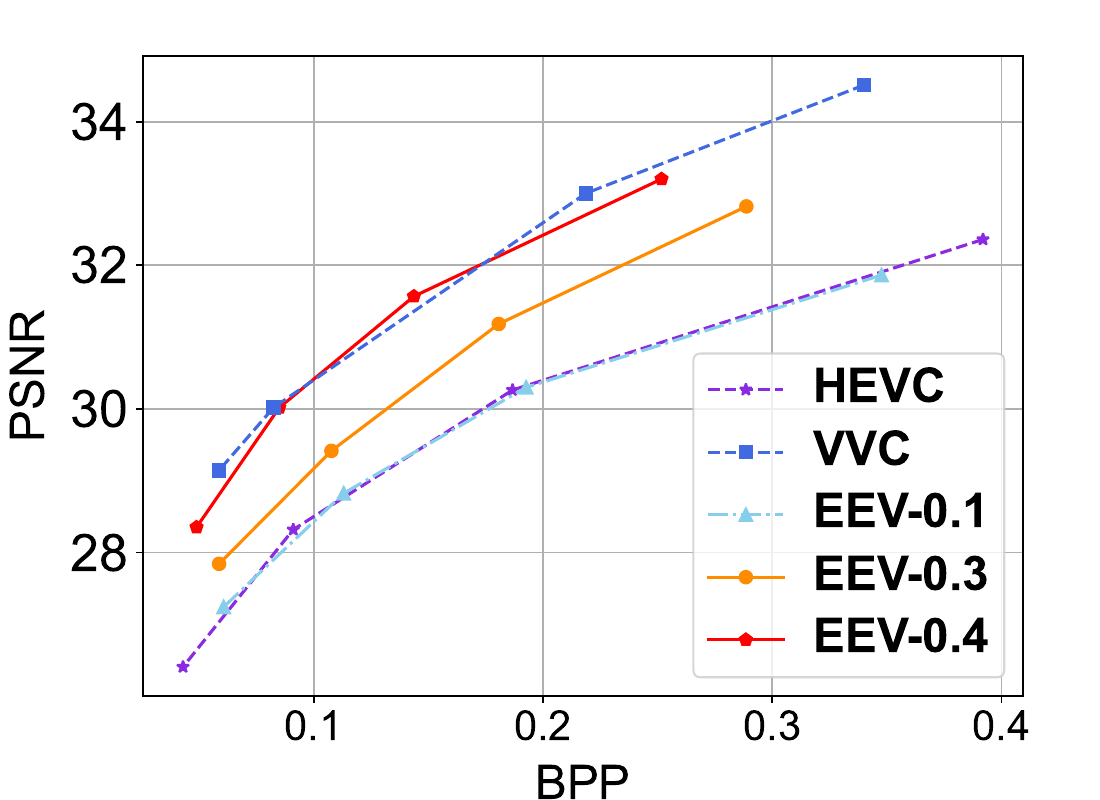}}
\caption{The R-D curves of the four UAV test sequences using PSNR metric. The performances of HEVC, VVC, EEV-0.1, EEV-0.3 and EEV-0.4 are depicted.}
\label{fig:rdcurves_psnr}
\end{figure*}

{\bf PSNR Metric.} We present the R-D curves of different coding methods using PSNR metric in Fig.~\ref{fig:rdcurves_psnr} and the detailed experimental results are available in Table.~\ref{tab:rdperformance-psnr}. Compared with traditional HEVC, EEV-0.3 and EEV-0.4 still dominate most of the data. EEV-0.3 tends to have better performance at high bit rate coding scenarios. However, on ClassC, which comprises indoor scenes, EEV's performance is sometimes inferior to HEVC. This could be due to the distortion and occlusion that are more noticeable when shooting indoor scenes with UAVs. Moreover, EEV-0.3 and EEV-0.4 are trained with Vimeo-90K, which lacks sufficient training samples for indoor scenes, leading to poor results when evaluating on such video sequences. Nevertheless, EEV-0.3 and EEV-0.4 exhibit superior performance in the other three categories.

\begin{figure*}[!htb]
    \centering
	\subfigure[PeopleOnStreet]{
	\includegraphics[width=0.23\textwidth]{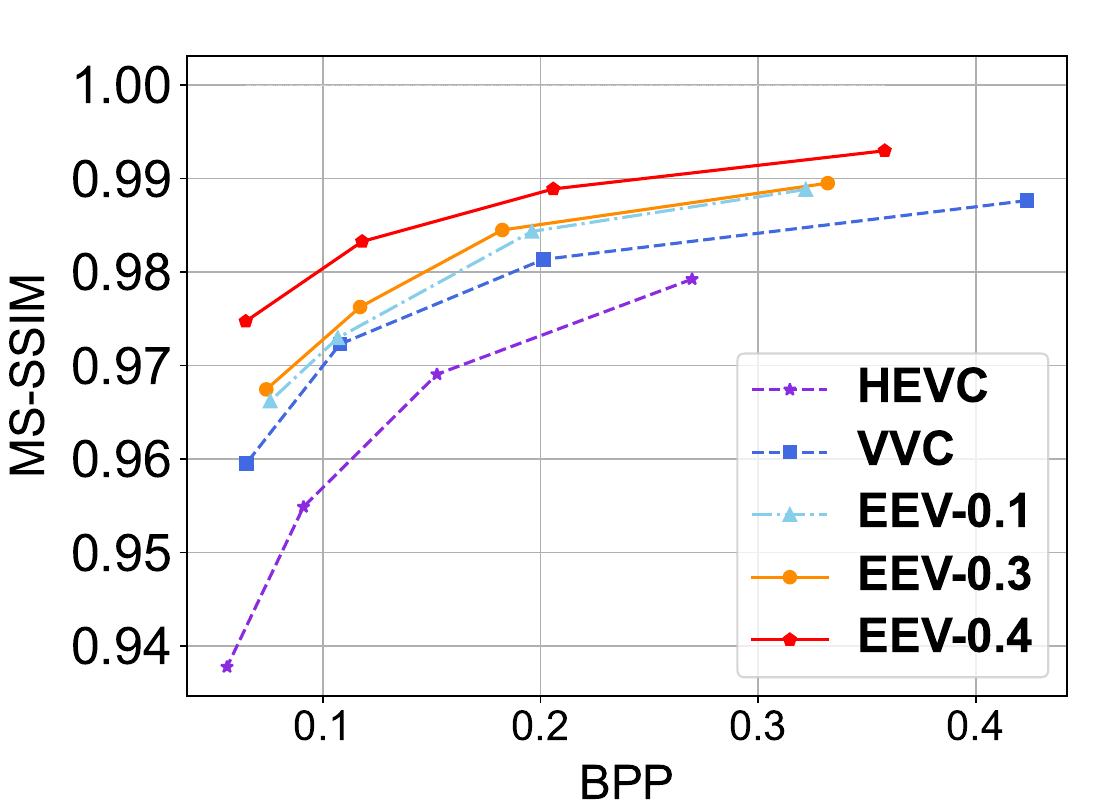}}
	\subfigure[Kimono]{
	\includegraphics[width=0.23\textwidth]{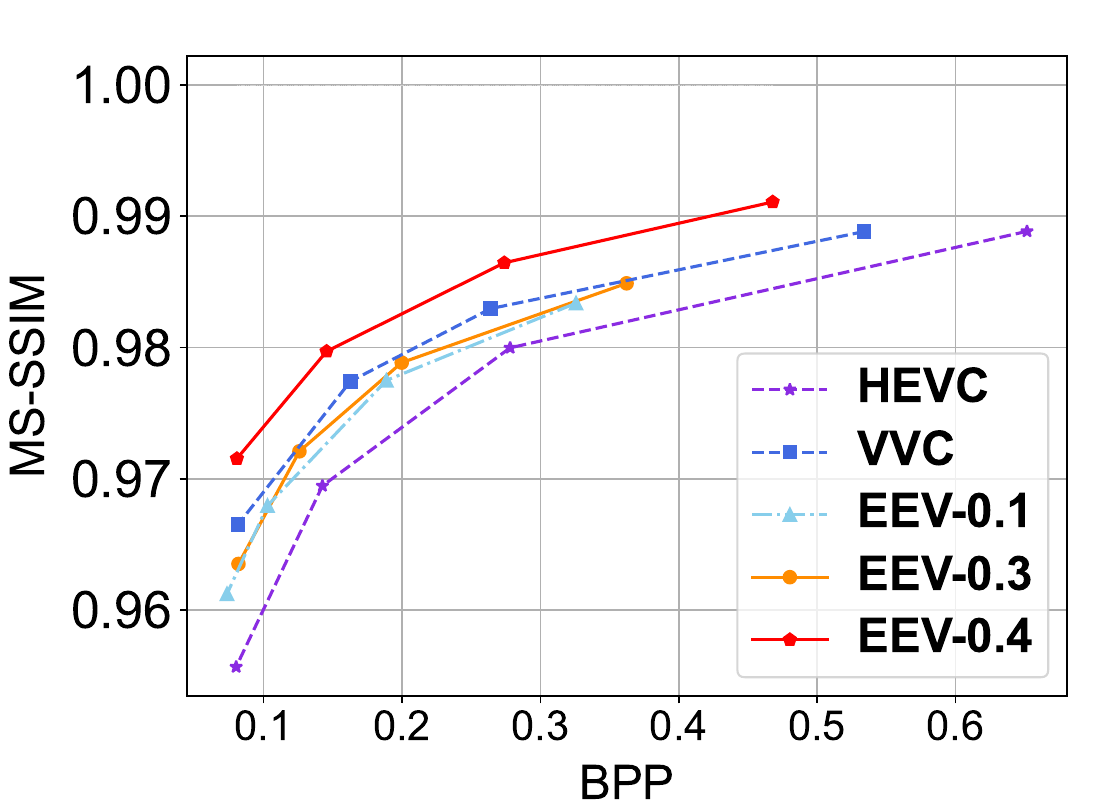}}
	\subfigure[PartyScene]{
	\includegraphics[width=0.23\textwidth]{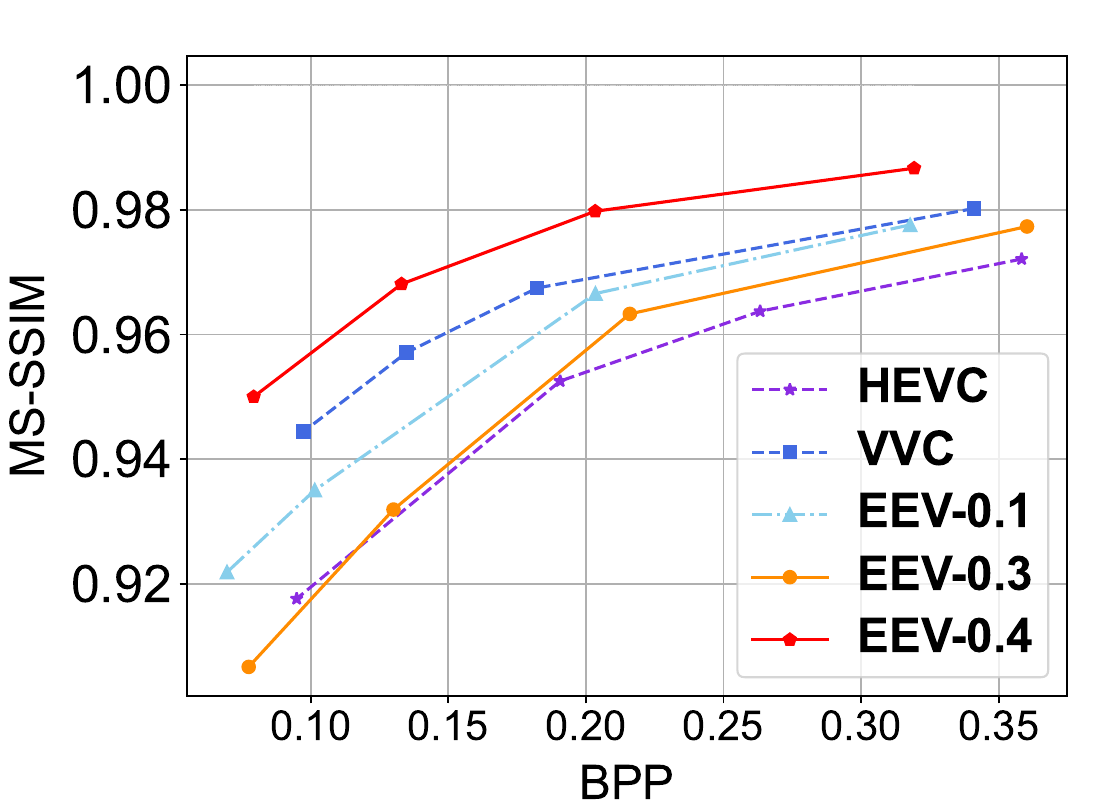}}
	\subfigure[FourPeople]{
	\includegraphics[width=0.23\textwidth]{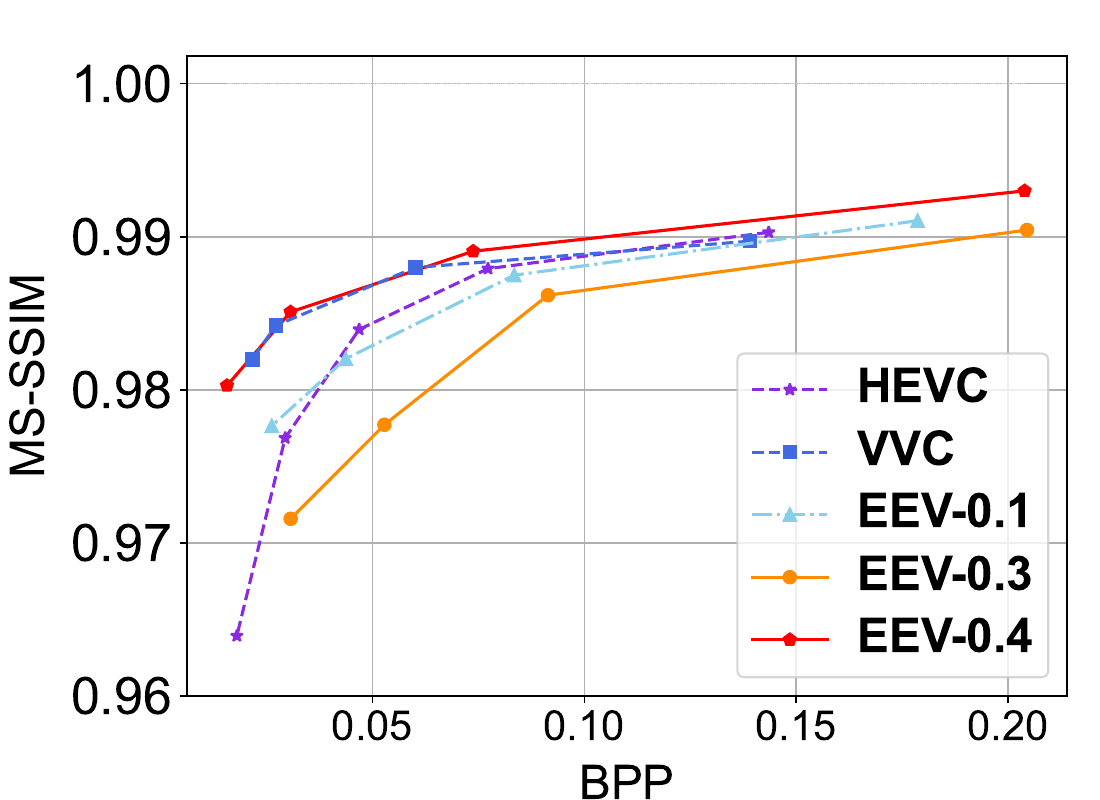}}
\caption{The R-D curves of the HEVC common test dataset using MS-SSIM metric. The performances of VVC, EEV-0.3 and EEV-0.4 are depicted. It is observed that EEV model has superior perceptual-quality oriented compression efficiency and outperforms the conventional coding scheme VVC.}
\label{fig:rdcurves_msssim_HEVC}
\end{figure*}

\begin{figure*}[!htb]
    \centering
	\subfigure[PeopleOnStreet]{
	\includegraphics[width=0.23\textwidth]{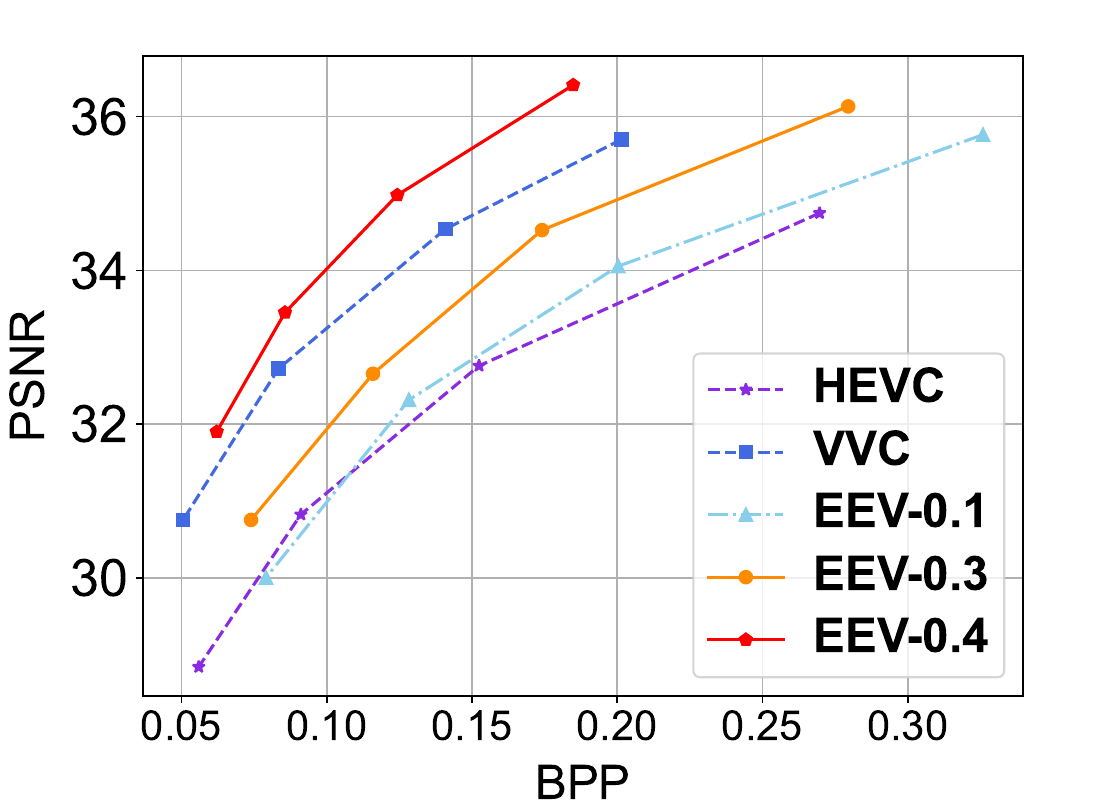}}
	\subfigure[Kimono]{
	\includegraphics[width=0.23\textwidth]{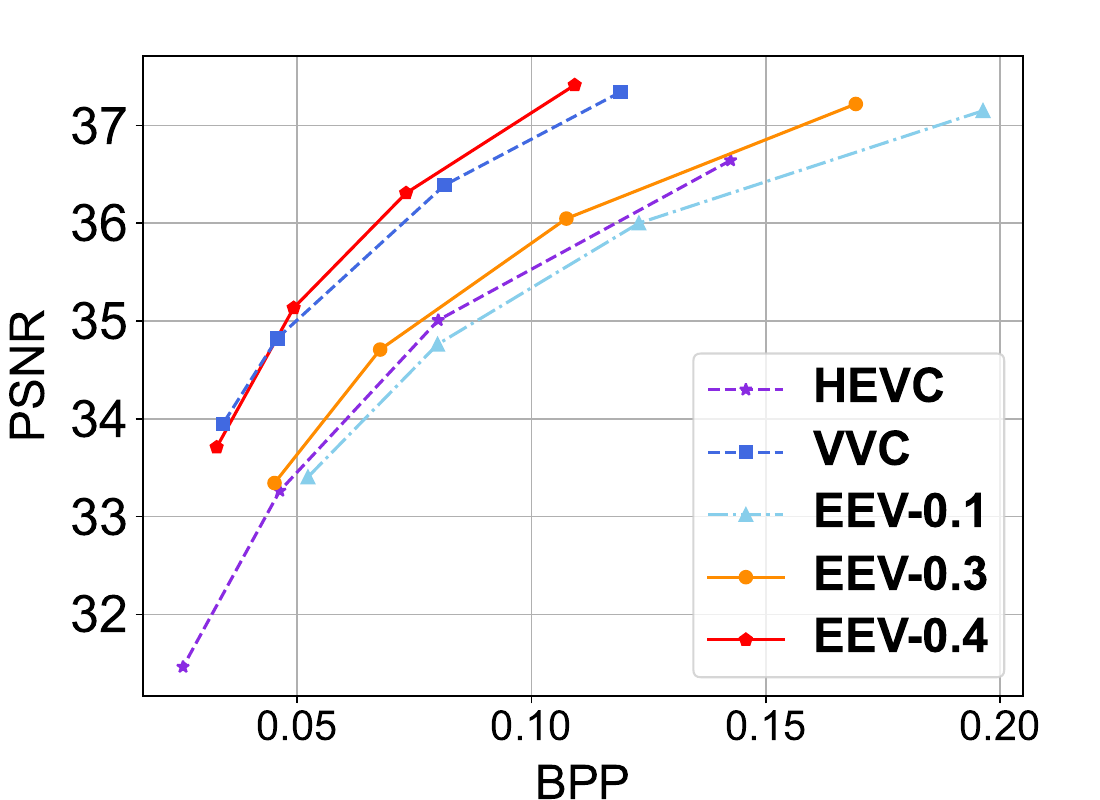}}
	\subfigure[PartyScene]{
	\includegraphics[width=0.23\textwidth]{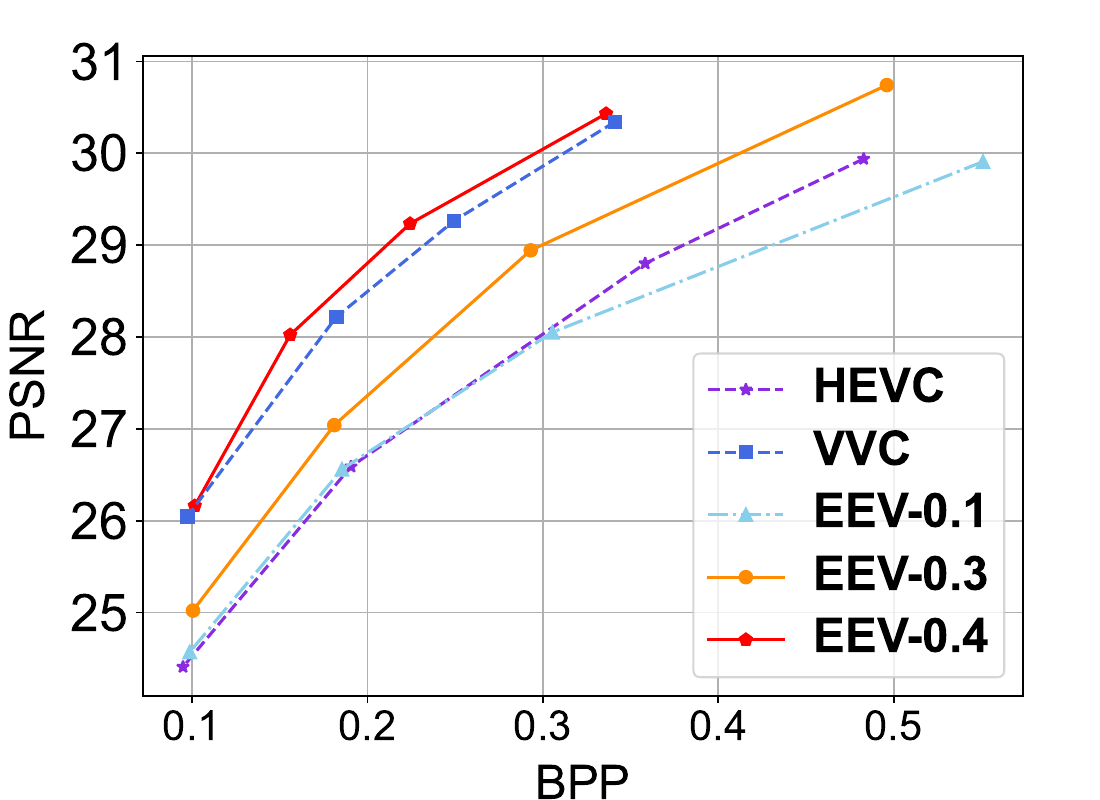}}
	\subfigure[FourPeople]{
	\includegraphics[width=0.23\textwidth]{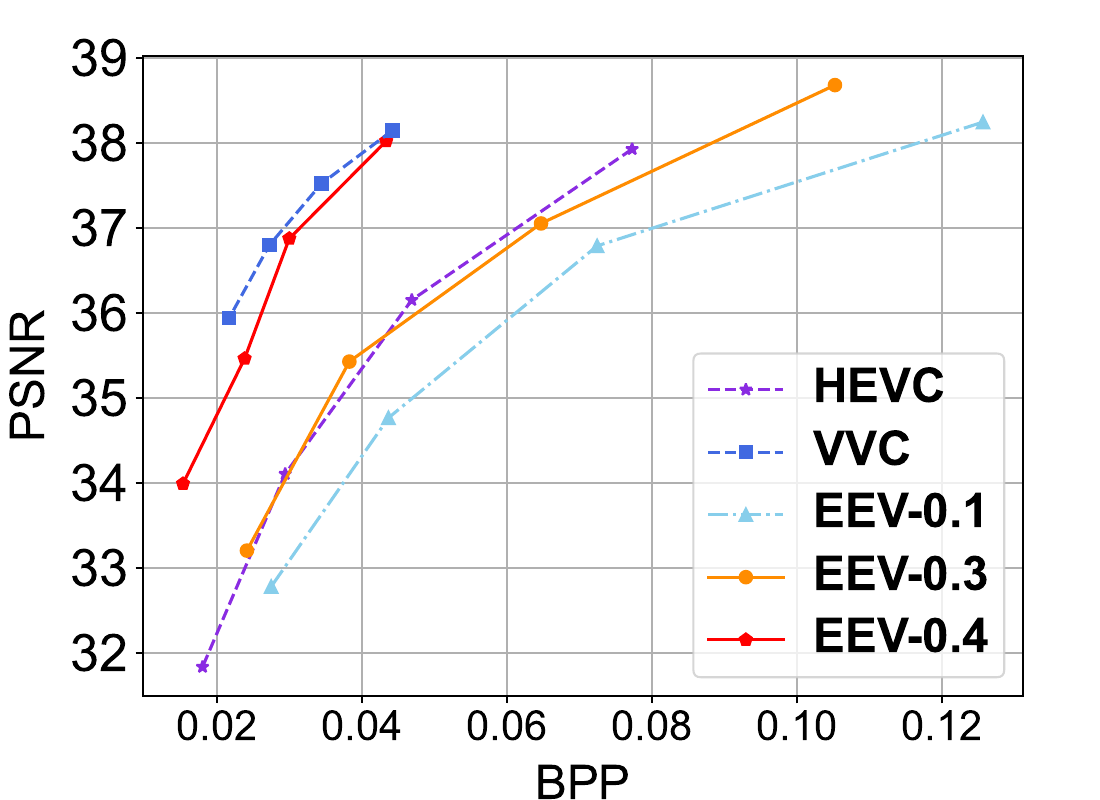}}
\caption{The R-D curves of the HEVC common test dataset using PSNR metric. The performances of VVC, EEV-0.3 and EEV-0.4 are depicted. It could be concluded that there exists clear space in pixel-fidelity oriented compression for EEV. }
\label{fig:rdcurves_psnr_HEVC}
\end{figure*}

{\bf Additional Tests.} In addition to testing the performance of VVC, EEV-0.3, and EEV-0.4 on the UAV sequences, we further evaluated the R-D efficiency on the HEVC dataset. The results, shown in Fig.~\ref{fig:rdcurves_msssim_HEVC} and Fig.~\ref{fig:rdcurves_psnr_HEVC}, indicate that the performance of each model is consistent with its performance on the UAV sequence. Regarding MS-SSIM metric, EEV-0.4 outperformed VVC in all categories, such as PeopleOnStreet, Kimono and PartyScene sequences. In most cases, the effect of EEV-0.3 was slightly inferior to that of VVC, but sometimes would surpass VVC such as in PeopleOnStreet sequence. For PSNR metric, EEV-0.4 performed better than VVC most of the time. The results for EEV-0.3 were inferior to VVC, but were obviously superior to HEVC in some sequences such as PeopleOnStreet and PartyScene. 

\begin{table*}[!htb]
\centering
\footnotesize
\caption{The BD-rate performance of different codecs (EEV-0.4, EEV-0.3, VTM-15.2 and HM-16.20-SCM-8.8) on HEVC CTC sequences and UVG dataset. The distortion metric is RGB-MSSSIM.}
\begin{tabular}{cc|c|c|c|c}
\hline
\multicolumn{1}{c|}{Category}                 & \begin{tabular}[c]{@{}c@{}}Sequence\\ Name\end{tabular} & \begin{tabular}[c]{@{}c@{}}BD-Rate Reduction\\ EEV-0.4 vs HEVC\end{tabular} & \begin{tabular}[c]{@{}c@{}}BD-Rate Reduction\\ EEV-0.3 vs HEVC\end{tabular} & \begin{tabular}[c]{@{}c@{}}BD-Rate Reduction\\ VVC vs HEVC\end{tabular} & \begin{tabular}[c]{@{}c@{}}BD-Rate Reduction\\ EEV-0.1 vs HEVC\end{tabular} \\ \hline
\multicolumn{1}{c|}{\multirow{2}{*}{Class A}} & Traffic                                                 & -42.87\%                                                                    & 21.60\%                                                                     & -34.64\%                                                                & -20.59\%                                                                    \\
\multicolumn{1}{c|}{}                         & PeopleOnStreet                                          & -68.40\%                                                                    & -47.50\%                                                                    & -39.55\%                                                                & -43.35\%                                                                    \\ \hline
\multicolumn{1}{c|}{\multirow{5}{*}{Class B}} & BasketballDrive                                         & -42.99\%                                                                    & 6.66\%                                                                      & -39.60\%                                                                & 11.04\%                                                                     \\
\multicolumn{1}{c|}{}                         & BQTerrace                                               & -52.26\%                                                                    & 22.77\%                                                                     & -43.57\%                                                                & -4.68\%                                                                     \\
\multicolumn{1}{c|}{}                         & ParkScene                                               & -32.58\%                                                                    & 7.20\%                                                                      & -31.24\%                                                                & -10.46\%                                                                    \\
\multicolumn{1}{c|}{}                         & Cactus                                                  & -44.90\%                                                                    & 11.37\%                                                                     & -33.61\%                                                                & -9.96\%                                                                     \\
\multicolumn{1}{c|}{}                         & Kimono                                                  & -48.17\%                                                                    & -24.56\%                                                                    & -33.90\%                                                                & -23.77\%                                                                    \\ \hline
\multicolumn{1}{c|}{\multirow{4}{*}{Class C}} & BasketballDrill                                         & -55.52\%                                                                    & 69.36\%                                                                     & -30.62\%                                                                & -12.61\%                                                                    \\
\multicolumn{1}{c|}{}                         & BQMALL                                                  & -54.35\%                                                                    & -4.26\%                                                                     & -43.18\%                                                                & -9.26\%                                                                     \\
\multicolumn{1}{c|}{}                         & PartyScene                                              & -57.57\%                                                                    & -4.86\%                                                                     & -37.90\%                                                                & -25.17\%                                                                    \\
\multicolumn{1}{c|}{}                         & RaceHorsesC                                             & -56.92\%                                                                    & -11.62\%                                                                    & -33.65\%                                                                & -3.87\%                                                                     \\ \hline
\multicolumn{1}{c|}{\multirow{4}{*}{Class D}} & BasketballPass                                          & -69.63\%                                                                    & -42.89\%                                                                    & -34.88\%                                                                & -47.06\%                                                                    \\
\multicolumn{1}{c|}{}                         & BlowingBubbles                                          & -63.09\%                                                                    & 11.44\%                                                                     & -35.36\%                                                                & -30.03\%                                                                    \\
\multicolumn{1}{c|}{}                         & BQSquare                                                & -74.41\%                                                                    & -38.58\%                                                                     & -38.13\%                                                                & -51.00\%                                                                    \\
\multicolumn{1}{c|}{}                         & RaceHorses                                              & -63.22\%                                                                    & -27.25\%                                                                    & -28.70\%                                                                & -20.92\%                                                                    \\ \hline
\multicolumn{1}{c|}{\multirow{3}{*}{Class E}} & FourPeople                                              & -45.06\%                                                                    & 60.28\%                                                                     & -40.02\%                                                                & 6.77\%                                                                      \\
\multicolumn{1}{c|}{}                         & Johnny                                                  & -29.22\%                                                                    & 168.76\%                                                                    & -43.28\%                                                                & 61.84\%                                                                     \\
\multicolumn{1}{c|}{}                         & KristenAndSara                                          & -47.38\%                                                                    & 54.18\%                                                                     & -41.04\%                                                                & 3.04\%                                                                      \\ \hline
\multicolumn{1}{c|}{\multirow{4}{*}{Class F}} & BasketballDrillText                                     & -47.95\%                                                                    & 79.88\%                                                                     & -31.92\%                                                                & -1.89\%                                                                     \\
\multicolumn{1}{c|}{}                         & ChinaSpeed                                              & -58.66\%                                                                    & -23.24\%                                                                    & -48.38\%                                                                & -27.95\%                                                                    \\
\multicolumn{1}{c|}{}                         & SlideEditing                                            & -34.58\%                                                                    & 123.84\%                                                                    & -63.34\%                                                                & 69.65\%                                                                     \\
\multicolumn{1}{c|}{}                         & SlideShow                                               & 55.89\%                                                                     & 523.50\%                                                                    & -55.06\%                                                                & 597.93\%                                                                    \\ \hline
\multicolumn{1}{c|}{\multirow{7}{*}{UVG}}     & Beauty                                                  & -33.38\%                                                                    & -6.38\%                                                                     & -33.21\%                                                                & -21.73\%                                                                    \\
\multicolumn{1}{c|}{}                         & Bosphorus                                               & -42.27\%                                                                    & 28.47\%                                                                     & -36.43\%                                                                & -6.08\%                                                                     \\
\multicolumn{1}{c|}{}                         & HoneyBee                                                & 18.20\%                                                                     & 191.57\%                                                                    & -11.81\%                                                                & 164.84\%                                                                    \\
\multicolumn{1}{c|}{}                         & Jockey                                                  & 9.08\%                                                                      & 162.79\%                                                                    & -33.78\%                                                                & 180.63\%                                                                    \\
\multicolumn{1}{c|}{}                         & ShakeNDry                                               & -42.94\%                                                                    & 0.06\%                                                                      & -41.35\%                                                                & -7.06\%                                                                     \\
\multicolumn{1}{c|}{}                         & YachtRide                                               & -68.34\%                                                                    & -36.96\%                                                                    & -39.82\%                                                                & -37.81\%                                                                     \\
\multicolumn{1}{c|}{}                         & ReadySteadyGo                                           & -19.22\%                                                                    & 25.84\%                                                                     & -34.57\%                                                                & 27.96\%                                                                     \\ \hline
\multicolumn{2}{c|}{Class A}                                                                            & -55.63\%                                                                    & -12.95\%                                                                    & -37.10\%                                                                & -31.97\%                                                                    \\
\multicolumn{2}{c|}{Class B}                                                                            & -44.18\%                                                                    & 4.69\%                                                                      & -36.38\%                                                                & -7.57\%                                                                     \\
\multicolumn{2}{c|}{Class C}                                                                            & -56.09\%                                                                    & 12.16\%                                                                     & -36.34\%                                                                & -12.73\%                                                                    \\
\multicolumn{2}{c|}{Class D}                                                                            & -67.59\%                                                                    & -24.32\%                                                                    & -34.27\%                                                                & -37.25\%                                                                    \\
\multicolumn{2}{c|}{Class E}                                                                            & -40.55\%                                                                    & 94.40\%                                                                     & -41.45\%                                                                & 23.88\%                                                                     \\
\multicolumn{2}{c|}{Class F}                                                                            & -21.33\%                                                                    & 175.99\%                                                                    & -49.68\%                                                                & 159.43\%                                                                    \\
\multicolumn{2}{c|}{UVG}                                                                                & -25.55\%                                                                    & 52.20\%                                                                     & -32.99\%                                                                & 42.96\%                                                                            \\ \hline
\multicolumn{2}{c|}{Average (Class A-E)}                                                                & -52.81\%                                                                    & 14.80\%                                                                     & -37.11\%                                                                & -13.13\%                                                                    \\ \hline
\multicolumn{2}{c|}{Overall Average}                                                                            & -44.42\%                                                                    & 43.17\%                                                                     & -38.31\%                                                                & 19.54\%                                                                    \\ \hline
\end{tabular}\label{tab:ctc-RGB-MSSSIM}
\end{table*}

\begin{figure}[!htb]
    \centering
	\subfigure[EEV-0.3 Computational Complexity]{
	\includegraphics[width=0.48\textwidth]{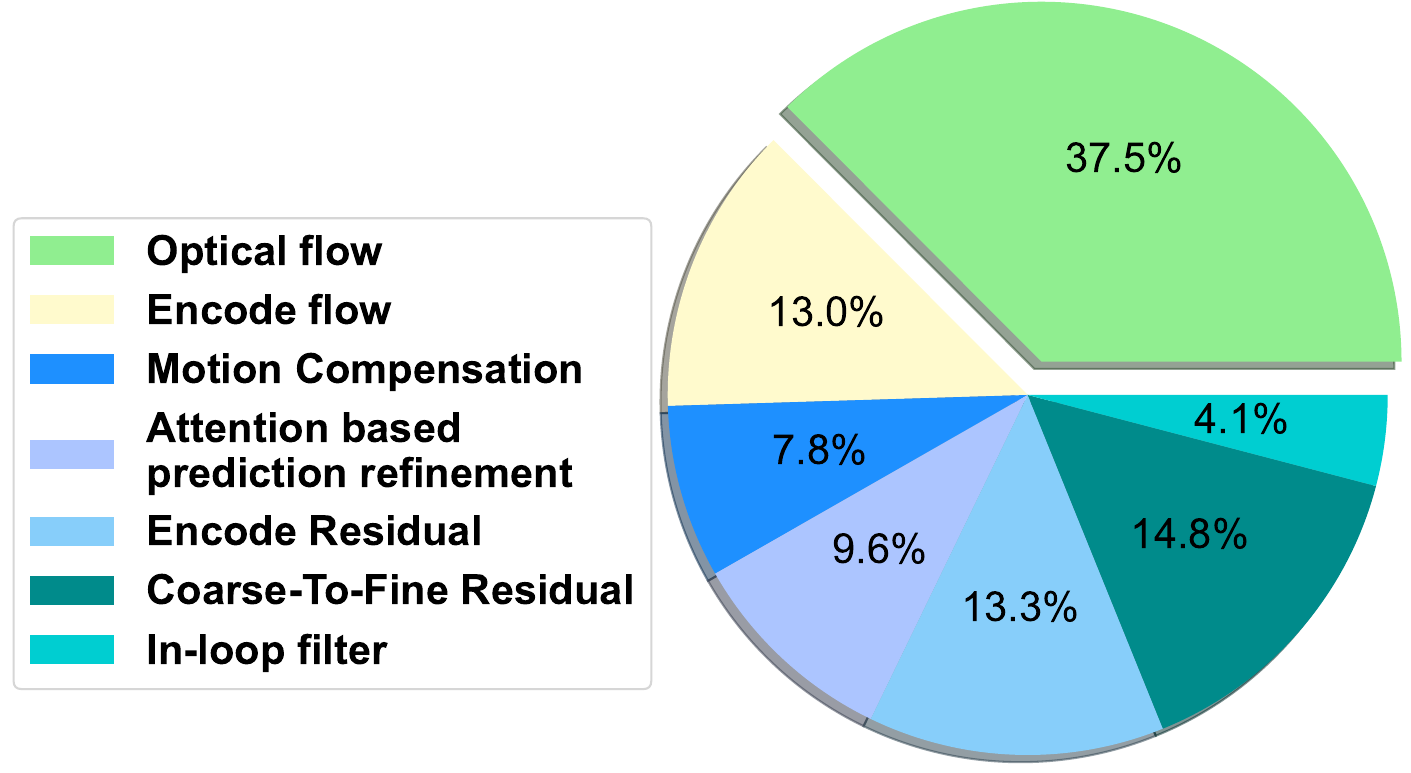}
	\label{PT(a)}}
	\subfigure[EEV-0.4 Computational Complexity]{
	\includegraphics[width=0.48\textwidth]{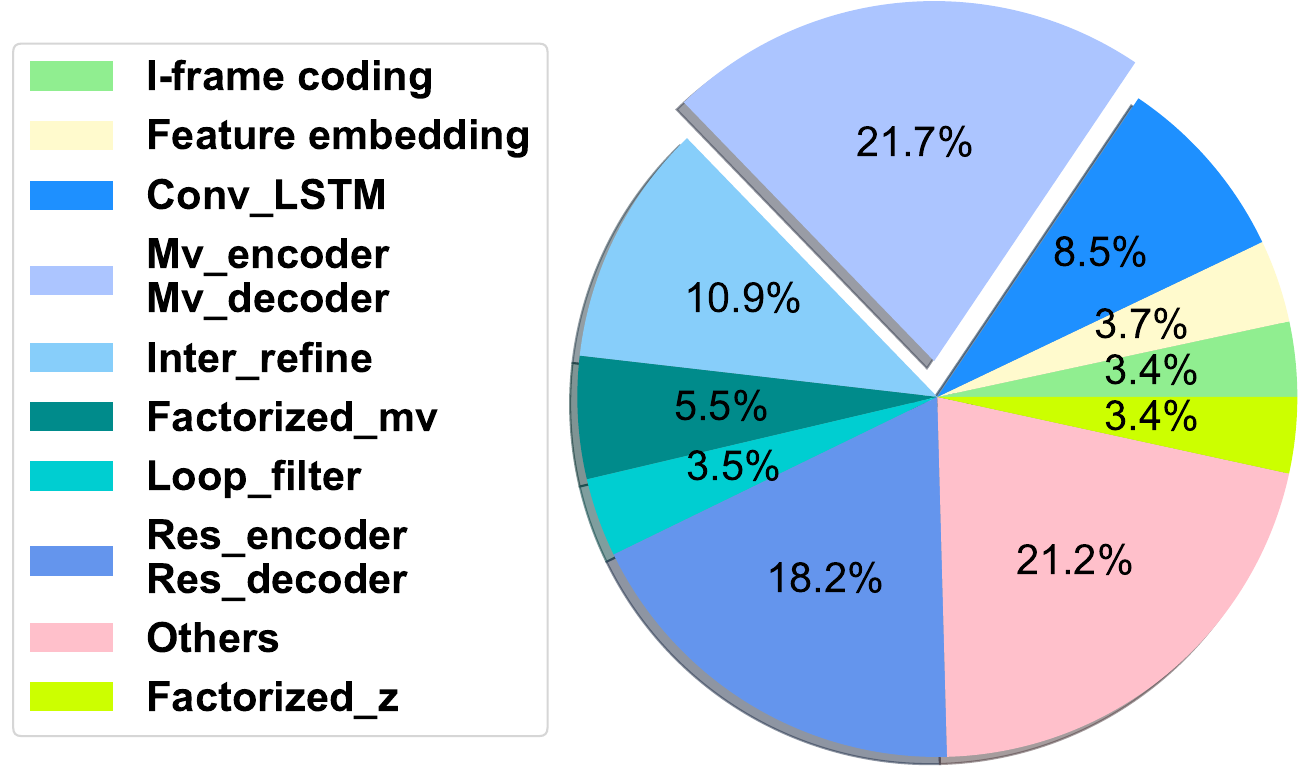}
	\label{PT(b)}}
    \caption{Run time complexity comparisons for different modules of EEV-0.3 and EEV-0.4. The above results are tested on a personal computer with Intel(R) Processor and NVIDIA GTX 1080Ti GPU.}
    \label{fig:Time}
\end{figure}

\begin{figure}[]
    \centering
	\subfigure[EEV-0.4 FLOPs Number]{
	\includegraphics[width=0.45\textwidth]{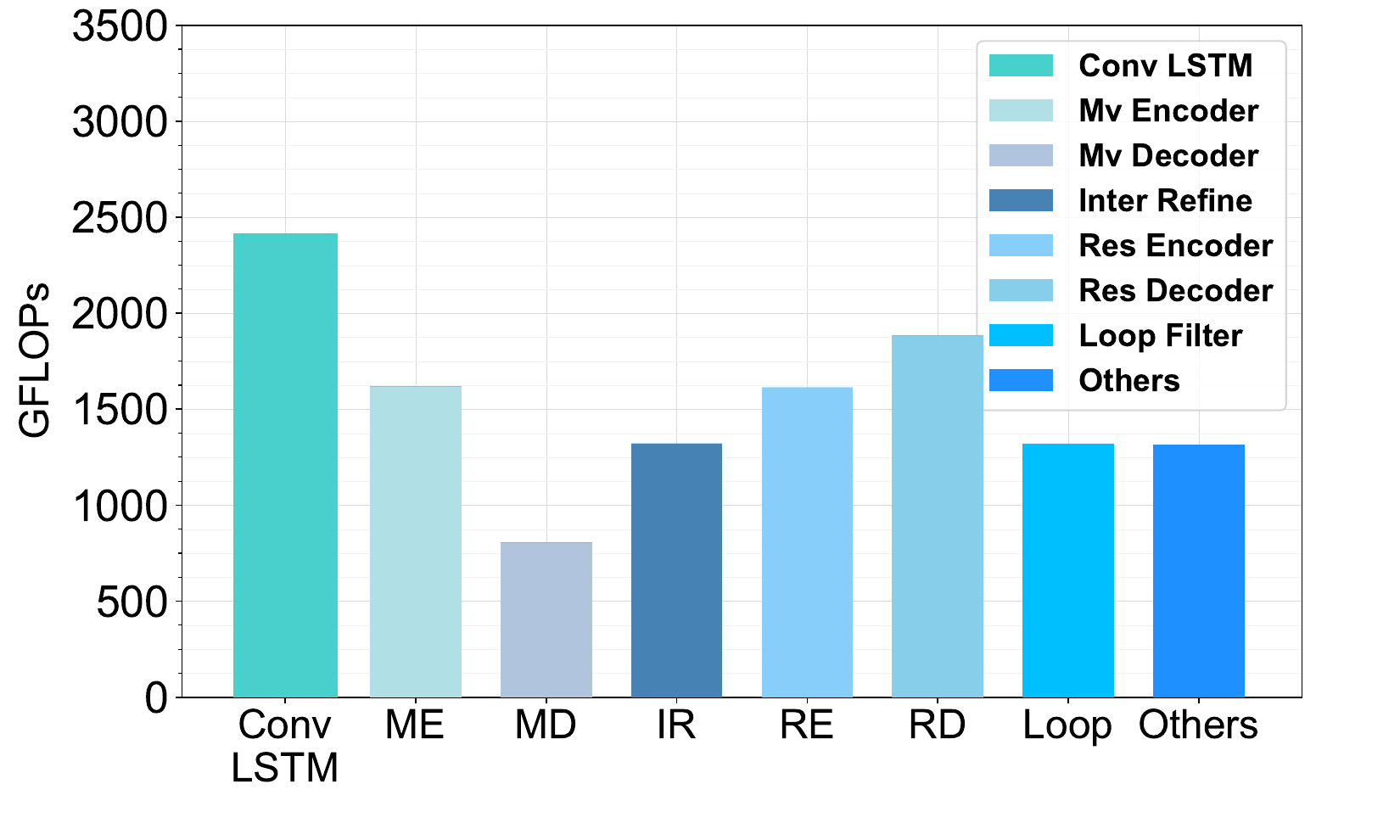}}
	\subfigure[EEV-0.3 FLOPs Number]{
	\includegraphics[width=0.45\textwidth]{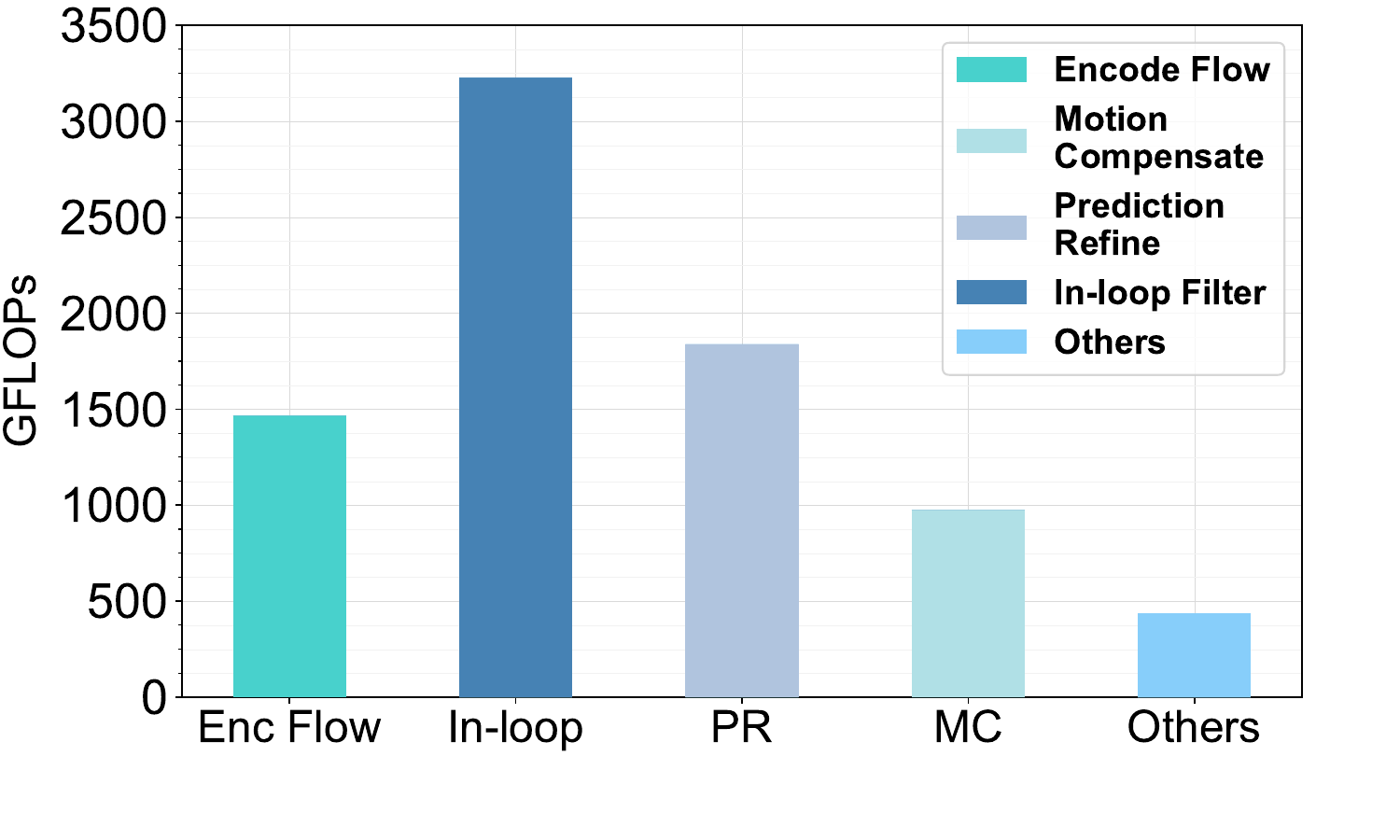}}
	\subfigure[EEV-0.4 Parameter Number]{
	\includegraphics[width=0.45\textwidth]{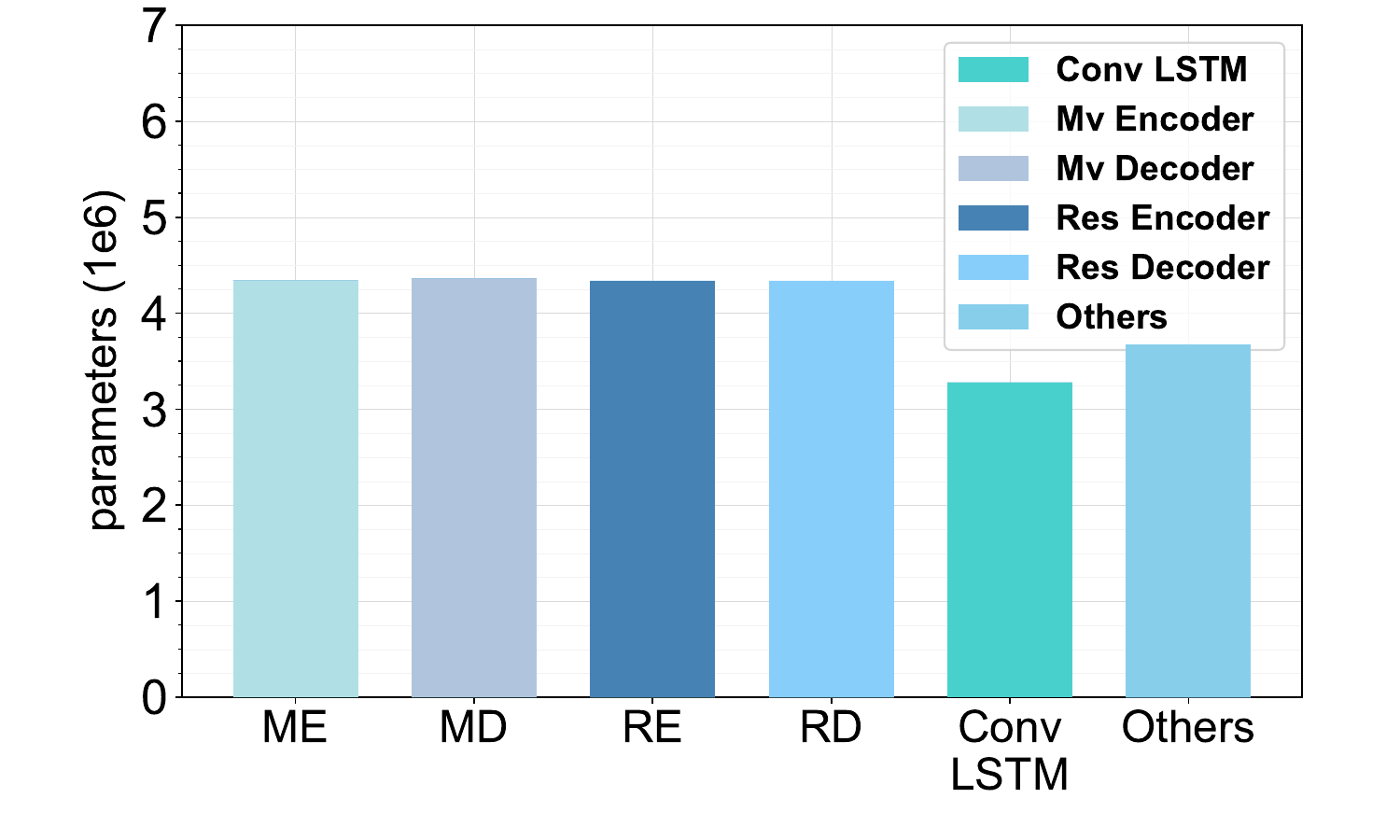}}
	\subfigure[EEV-0.3 Parameter Number]{
	\includegraphics[width=0.45\textwidth]{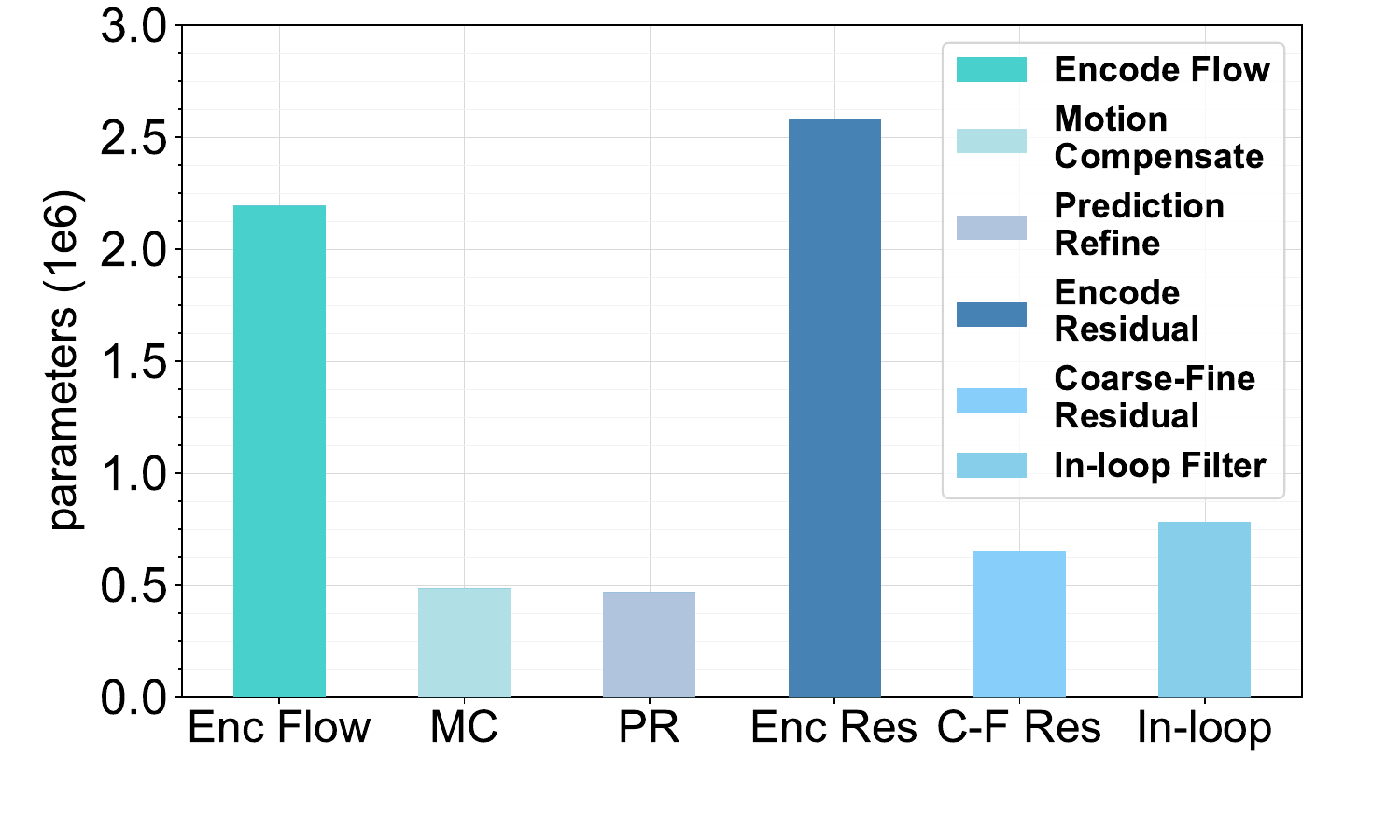}}
\caption{The FLOPs and the total number of learnable parameters of EEV-0.3 and EEV-0.4 model. The $ME$, $MD$, $IR$, $RE$, $RD$, $Loop$ denote MV encoder, MV decoder, inter refinement, residual encoder, residual decoder, and in-loop filter network respectively.}
\label{fig:FLOPs}
\end{figure}

\begin{table*}[]
\centering
\footnotesize
\caption{The BD-rate performance of different codecs (EEV-0.4, EEV-0.3, VTM-15.2 and HM-16.20-SCM-8.8) on HEVC CTC sequences and UVG dataset. The distortion metric is RGB-PSNR.}
\begin{tabular}{cc|c|c|c|c}
\hline
\multicolumn{1}{c|}{\multirow{2}{*}{Category}} & \multirow{2}{*}{\begin{tabular}[c]{@{}c@{}}Sequence\\ Name\end{tabular}} & \multirow{2}{*}{\begin{tabular}[c]{@{}c@{}}BD-Rate Reduction\\ EEV-0.4 vs HEVC\end{tabular}} & \multirow{2}{*}{\begin{tabular}[c]{@{}c@{}}BD-Rate Reduction\\ EEV-0.3 vs HEVC\end{tabular}} & \multirow{2}{*}{\begin{tabular}[c]{@{}c@{}}BD-Rate Reduction\\ VVC vs HEVC\end{tabular}} & \multirow{2}{*}{\begin{tabular}[c]{@{}c@{}}BD-Rate Reduction\\ EEV-0.1 vs HEVC\end{tabular}} \\
\multicolumn{1}{c|}{}                          &                                                                          &                                                                                              &                                                                                              &                                                                                          &                                                                                              \\ \hline
\multicolumn{1}{c|}{\multirow{2}{*}{Class A}}  & Traffic                                                                  & -32.94\%                                                                                     & -0.87\%                                                                                      & -35.34\%                                                                                 & 25.69\%                                                                                      \\
\multicolumn{1}{c|}{}                          & PeopleOnStreet                                                           & -53.15\%                                                                                     & -23.37\%                                                                                     & -44.43\%                                                                                 & -4.01\%                                                                                      \\ \hline
\multicolumn{1}{c|}{\multirow{5}{*}{Class B}}  & BasketballDrive                                                          & -9.63\%                                                                                      & 59.81\%                                                                                      & -44.08\%                                                                                 & 105.98\%                                                                                     \\
\multicolumn{1}{c|}{}                          & BQTerrace                                                                & -41.09\%                                                                                     & 8.77\%                                                                                       & -42.72\%                                                                                 & 48.07\%                                                                                      \\
\multicolumn{1}{c|}{}                          & ParkScene                                                                & -37.38\%                                                                                     & -10.44\%                                                                                     & -36.08\%                                                                                 & 8.03\%                                                                                       \\
\multicolumn{1}{c|}{}                          & Cactus                                                                   & -43.01\%                                                                                     & -3.61\%                                                                                      & -41.66\%                                                                                 & 26.88\%                                                                                      \\
\multicolumn{1}{c|}{}                          & Kimono                                                                   & -40.78\%                                                                                     & -6.41\%                                                                                      & -38.68\%                                                                                 & 8.43\%                                                                                       \\ \hline
\multicolumn{1}{c|}{\multirow{4}{*}{Class C}}  & BasketballDrill                                                          & -28.55\%                                                                                     & 11.05\%                                                                                      & -33.60\%                                                                                 & 47.13\%                                                                                      \\
\multicolumn{1}{c|}{}                          & BQMALL                                                                   & -36.72\%                                                                                     & 16.08\%                                                                                      & -43.85\%                                                                                 & 49.57\%                                                                                      \\
\multicolumn{1}{c|}{}                          & PartyScene                                                               & -43.74\%                                                                                     & -18.05\%                                                                                     & -39.53\%                                                                                 & 2.59\%                                                                                       \\
\multicolumn{1}{c|}{}                          & RaceHorsesC                                                              & -27.50\%                                                                                     & 16.79\%                                                                                      & -36.04\%                                                                                 & 53.25\%                                                                                      \\ \hline
\multicolumn{1}{c|}{\multirow{4}{*}{Class D}}  & BasketballPass                                                           & -46.69\%                                                                                     & -25.75\%                                                                                     & -35.77\%                                                                                 & -1.17\%                                                                                      \\
\multicolumn{1}{c|}{}                          & BlowingBubbles                                                           & -47.47\%                                                                                     & -11.40\%                                                                                     & -35.35\%                                                                                 & 10.77\%                                                                                      \\
\multicolumn{1}{c|}{}                          & BQSquare                                                                 & -57.25\%                                                                                     & -39.83\%                                                                                     & -39.39\%                                                                                 & -22.06\%                                                                                     \\
\multicolumn{1}{c|}{}                          & RaceHorses                                                               & -45.74\%                                                                                     & -20.56\%                                                                                     & -33.87\%                                                                                 & 5.37\%                                                                                       \\ \hline
\multicolumn{1}{c|}{\multirow{3}{*}{Class E}}  & FourPeople                                                               & -44.10\%                                                                                     & 0.87\%                                                                                       & -50.51\%                                                                                 & 29.54\%                                                                                      \\
\multicolumn{1}{c|}{}                          & Johnny                                                                   & -31.55\%                                                                                     & 47.74\%                                                                                      & -47.72\%                                                                                 & 85.31\%                                                                                      \\
\multicolumn{1}{c|}{}                          & KristenAndSara                                                           & -36.96\%                                                                                     & 30.74\%                                                                                      & -48.91\%                                                                                 & 74.08\%                                                                                      \\ \hline
\multicolumn{1}{c|}{\multirow{4}{*}{Class F}}  & BasketballDrillText                                                      & -10.39\%                                                                                     & 43.50\%                                                                                      & -35.46\%                                                                                 & 88.56\%                                                                                      \\
\multicolumn{1}{c|}{}                          & ChinaSpeed                                                               & 80.27\%                                                                                      & 150.24\%                                                                                     & -46.21\%                                                                                 & 200.65\%                                                                                     \\
\multicolumn{1}{c|}{}                          & SlideEditing                                                             & 61.43\%                                                                                      & 118.75\%                                                                                     & -61.79\%                                                                                 & 142.42\%                                                                                     \\
\multicolumn{1}{c|}{}                          & SlideShow                                                                & 163.81\%                                                                                     & 362.47\%                                                                                     & -55.79\%                                                                                 & 470.24\%                                                                                     \\ \hline
\multicolumn{1}{c|}{\multirow{7}{*}{UVG}}      & Beauty                                                                   & -61.06\%                                                                                     & -7.66\%                                                                                      & -48.30\%                                                                                 & -14.37\%                                                                                     \\
\multicolumn{1}{c|}{}                          & Bosphorus                                                                & -25.15\%                                                                                     & 40.40\%                                                                                      & -42.01\%                                                                                 & 48.04\%                                                                                      \\
\multicolumn{1}{c|}{}                          & HoneyBee                                                                 & -1.57\%                                                                                      & 88.16\%                                                                                      & -21.46\%                                                                                 & 150.71\%                                                                                     \\
\multicolumn{1}{c|}{}                          & Jockey                                                                   & 103.64\%                                                                                     & 294.07\%                                                                                     & -38.99\%                                                                                 & 381.03\%                                                                                     \\
\multicolumn{1}{c|}{}                          & ShakeNDry                                                                & -22.90\%                                                                                     & -4.32\%                                                                                      & -46.54\%                                                                                 & 13.75\%                                                                                      \\
\multicolumn{1}{c|}{}                          & YachtRide                                                                & -46.28\%                                                                                     & -11.23\%                                                                                     & -44.92\%                                                                                 & 18.26\%                                                                                      \\
\multicolumn{1}{c|}{}                          & ReadySteadyGo                                                            & 9.14\%                                                                                       & 38.80\%                                                                                      & -38.72\%                                                                                 & 91.17\%                                                                                      \\ \hline
\multicolumn{2}{c|}{Class A}                                                                                              & -43.04\%                                                                                     & -12.12\%                                                                                     & -39.89\%                                                                                 & 10.84\%                                                                                      \\
\multicolumn{2}{c|}{Class B}                                                                                              & -34.38\%                                                                                     & 9.63\%                                                                                       & -40.64\%                                                                                 & 39.48\%                                                                                      \\
\multicolumn{2}{c|}{Class C}                                                                                              & -34.13\%                                                                                     & 6.47\%                                                                                       & -38.26\%                                                                                 & 38.14\%                                                                                      \\
\multicolumn{2}{c|}{Class D}                                                                                              & -49.29\%                                                                                     & -24.38\%                                                                                     & -36.10\%                                                                                 & -1.77\%                                                                                      \\
\multicolumn{2}{c|}{Class E}                                                                                              & -37.54\%                                                                                     & 26.45\%                                                                                      & -49.05\%                                                                                 & 62.97\%                                                                                      \\
\multicolumn{2}{c|}{Class F}                                                                                              & 73.78\%                                                                                      & 168.74\%                                                                                     & -49.81\%                                                                                 & 225.47\%                                                                                     \\
\multicolumn{2}{c|}{UVG}                                                                                                  & -6.31\%                                                                                      & 62.60\%                                                                                      & -40.14\%                                                                                 & 98.37\%                                                                                      \\ \hline
\multicolumn{2}{c|}{Average (Class A-E)}                                                                                  & -39.68\%                                                                                     & 1.21\%                                                                                       & -40.79\%                                                                                 & 29.93\%                                                                                      \\ \hline
\multicolumn{2}{c|}{Overall Average}                                                                                              & -18.70\%                                                                                     & 33.91\%                                                                                      & -41.98\%                                                                                 & 67.64\%                                                                                      \\ \hline
\end{tabular}
\label{tab:ctc-PSNR}
\end{table*}

{\bf Model Complexity.} To further illustrate the model complexity as well as the computational cost of EEV reference models, the FLOPs and the total number of learnable parameters for EEV-0.3 and EEV-0.4 are shown in Fig.~\ref{fig:FLOPs}. For EEV-0.3, the modules with the highest number of calculations are the in-loop filter responsible for removing compression artifacts. The coding residual has the largest number of parameters. For EEV-0.4, the total number of operations is slightly higher than EEV-0.3, with the coding residual also being the main computational overhead. By comparative analysis, it is observed that the total number of EEV-0.4 parameters is significantly higher than that of EEV-0.3. The reason behind this phenomenon lies in that the motion and residual encoding and decoding require more parameters. This could be further optimized using entropy skipping or other methods. 

To evaluate the absolute inference time of EEV reference models, videos with the resolution of 1920$\times$1024 are adopted for evaluation. Specifically, we have collected the running time over the entire test sequence. The running time of PSNR- and MS-SSIM-optimized models is almost the same for EEV-0.4, which is about 185s for 100 frames, with most of the time spent on data communication between CPU and GPU, and model loading.

Regarding EEV-0.3, the execution time of PSNR-optimized model is around 260s, with most of the time spent on P frame compression, including optical flow calculation, motion compensation, and other modules. The running time of the MS-SSIM-optimized model is even longer, about 1680s, with most of the time spent on encoding and decoding I-frames. This is mainly because of the autoregressive entropy model when calculating the probability for arithmetic coding.

To investigate the bottleneck module of computational cost, we further present the time consumption of each module involved in compressing P frames, and the corresponding result is plotted in Fig.~\ref{fig:Time}. As for EEV-0.3, the most time-consuming task is computing optical flow, which represents over one-third of the total time. Encoding, decoding, and refining the residual also require a significant amount of time. Interestingly, the In-loop filter with higher FLOPs only takes up 4.1\% of the total time. For EEV-0.4, motion and residual encoding and decoding are the most time-consuming tasks, accounting for around 40\% of the total time, consistent with the FLOPs and parameter count results.

We additionally summarize the model parameters, bit-depth of model parameters, and MACs per pixel for each version of our model. The following Table~\ref{tab:model-complexity-macs} presents the comprehensive comparisons. It should be pointed out that the network of EEV-0.2 is a subset of EEV-0.3, therefore, we list the numbers of EEV-0.1, EEV-0.3, and EEV-0.4. In future design, an important factor is to reduce the model size and computational complexity for EEV models.

\begin{table}[!htb]
\centering
\caption{Model complexity of EEV models}
\begin{tabular}{c|cccc}
\hline
\textbf{Model} & \textbf{\begin{tabular}[c]{@{}c@{}}EEV-0.1\end{tabular}} & \textbf{\begin{tabular}[c]{@{}c@{}}EEV-0.3\end{tabular}} & \textbf{\begin{tabular}[c]{@{}c@{}}EEV-0.4\end{tabular}} \\ \hline
MACs per pixel (M)              & 0.678                                            & 2.021                                              & 3.127                                          \\
Parameters (M)              & 5.26                                            & 7.17                                              & 23.96                                          \\ 
Weights bit-depth              & FP32                                            & FP32                                              & FP32                                          \\
\hline 
\end{tabular}
\label{tab:model-complexity-macs}
\end{table}

Currently, the execution time of neural codecs is mainly dominated by the encoding and decoding of motion vectors, which are employed to predict the movements between adjacent frames. While in traditional codecs, inter-frame prediction results in heavy complexity, as it takes over 40\% of running time in searching corresponding moving patterns and blocks between frames~\cite{vanne2012comparative}. As such, traditional codecs make a rate-distortion-complexity joint optimization by using pruning or other methods in the searching procedure to reduce the complexity, which may also cause performance degradation in rate-distortion performance. However, in neural network codecs, the reduction of complexity is mainly achieved by model pruning (reducing the parameters and FLOPs in NNs), knowledge distillation (migrating the generalization from complex models to simple models), and so on, with the target of scaling down models for easier deployment while maintaining the capacity of models.

{\bf Different Quality Evaluation Metrics.} The PSNR and MS-SSIM values might not be able to reveal the compression performance from the pespective of subjective quality. We have conducted extensive experimental results by using VMAF as distortion metric. The test sequences are the drone videos used in MPAI-EEV group. VMAF is a widely employed quality metric by fusing several objective metrics in some standardization organizations. The following Table~\ref{tab:bd-rate-vmaf} depicts the BD-rate performances using VMAF (mse-optimized model). It is observed that the VVC codec has better compression performance than other codecs. The results also reveal that the learned video codecs such as EEV-0.4 have better coding performances against HEVC with clear margins. It should be pointed out that the BD-rate reduction performances are consistent among different classes, demonstrating that EEV models are robust and with generalization ability.

\begin{table*}[!htb]
\centering
\scriptsize
\caption{The BD-rate performance of different codecs (EEV-0.4, EEV-0.3, VTM-15.2 and HM-16.20-SCM-8.8) on drone video compression. The distortion metric is VMAF.}
\begin{tabular}{cc|c|c|c|c}
\hline
\multicolumn{1}{c|}{Category}                 & \begin{tabular}[c]{@{}c@{}}Sequence\\ Name\end{tabular} & \begin{tabular}[c]{@{}c@{}}BD-Rate Reduction\\ EEV-0.4 vs HEVC\end{tabular} & \begin{tabular}[c]{@{}c@{}}BD-Rate Reduction\\ EEV-0.3 vs HEVC\end{tabular} & \begin{tabular}[c]{@{}c@{}}BD-Rate Reduction\\ VVC vs HEVC\end{tabular} & \begin{tabular}[c]{@{}c@{}}BD-Rate Reduction\\ EEV-0.1 vs HEVC\end{tabular} \\ \hline
\multicolumn{1}{c|}{\multirow{5}{*}{Class A}} & BasketballGround                                        & -51.39\%                                                                    & -24.56\%                                                                    & -52.36\%                                                                & -19.82\%                                                                    \\
\multicolumn{1}{c|}{}                         & GrassLand                                               & -48.17\%                                                                    & -32.63\%                                                                    & -60.82\%                                                                & -42.52\%                                                                    \\
\multicolumn{1}{c|}{}                         & Intersection                                            & -57.68\%                                                                    & -34.37\%                                                                    & -60.20\%                                                                & -39.95\%                                                                    \\
\multicolumn{1}{c|}{}                         & NightMall                                               & -50.08\%                                                                    & -24.06\%                                                                    & -52.83\%                                                                & -18.98\%                                                                    \\
\multicolumn{1}{c|}{}                         & SoccerGround                                            & -45.55\%                                                                    & -20.25\%                                                                    & -71.18\%                                                                & -30.09\%                                                                    \\ \hline
\multicolumn{1}{c|}{\multirow{3}{*}{Class B}} & Circle                                                  & -47.30\%                                                                    & -37.37\%                                                                    & -59.12\%                                                                & -29.25\%                                                                    \\
\multicolumn{1}{c|}{}                         & CrossBridge                                             & -9.72\%                                                                     & 16.35\%                                                                     & -58.84\%                                                                & 69.03\%                                                                     \\
\multicolumn{1}{c|}{}                         & Highway                                                 & -44.46\%                                                                    & -13.65\%                                                                    & -53.16\%                                                                & 2.50\%                                                                      \\ \hline
\multicolumn{1}{c|}{\multirow{3}{*}{Class C}} & Classroom                                               & -37.66\%                                                                    & 25.71\%                                                                     & -47.93\%                                                                & 11.70\%                                                                     \\
\multicolumn{1}{c|}{}                         & Elevator                                                & -47.19\%                                                                    & -8.04\%                                                                     & -46.90\%                                                                & 7.31\%                                                                      \\
\multicolumn{1}{c|}{}                         & Hall                                                    & -52.61\%                                                                    & -14.78\%                                                                    & -52.72\%                                                                & -31.71\%                                                                    \\ \hline
\multicolumn{1}{c|}{\multirow{3}{*}{Class D}} & Campus                                                  & -54.58\%                                                                    & -35.27\%                                                                    & -53.30\%                                                                & -21.12\%                                                                    \\
\multicolumn{1}{c|}{}                         & RoadByTheSea                                            & -51.46\%                                                                    & -32.86\%                                                                    & -51.08\%                                                                & -19.43\%                                                                    \\
\multicolumn{1}{c|}{}                         & Theater                                                 & -26.77\%                                                                    & 4.85\%                                                                      & -56.38\%                                                                & 31.28\%                                                                     \\ \hline
\multicolumn{2}{c|}{Class A}                                                                            & -50.57\%                                                                    & -27.17\%                                                                    & -59.48\%                                                                & -30.27\%                                                                    \\
\multicolumn{2}{c|}{Class B}                                                                            & -33.83\%                                                                    & -11.55\%                                                                    & -57.04\%                                                                & 14.09\%                                                                     \\
\multicolumn{2}{c|}{Class C}                                                                            & -45.82\%                                                                    & 0.96\%                                                                      & -49.18\%                                                                & -4.23\%                                                                     \\
\multicolumn{2}{c|}{Class D}                                                                            & -44.27\%                                                                    & -21.09\%                                                                    & -53.59\%                                                                & -3.09\%                                                                     \\ \hline
\multicolumn{2}{c|}{Average}                                                                            & -44.62\%                                                                    & -16.49\%                                                                    & -55.49\%                                                                & -9.36\%                                                                     \\ \hline
\end{tabular}
\label{tab:bd-rate-vmaf}
\end{table*}

{\bf Possible Improvements.} Currently, both EEV-0.3 and EEV-0.4 exhibit poor performance in indoor scenes. This may be attributed to the severe deformation and distortion commonly found in indoor environments, or the models' lack of training on UAV indoor scene datasets. However, the limited availability of drone-captured indoor data poses a significant challenge for effective model training. To overcome this, it may be necessary to construct additional UAV video datasets, particularly for indoor scenes, to improve the effectiveness of model training in the future.

Furthermore, according to the R-D figures, it is observed that the default model in~\cite{begaint2020compressai} might not be as effective at compressing I-frames as BPG image codec\footnote{https://bellard.org/bpg} on certain sequences such as \textit{BasketballGround}, leading to a lower overall PSNR value. Hence, conducting additional training on intra-coding using the UAV dataset may be a promising solution for improvement.

\section{Open Discussions and Future Remarks}\label{discuss}
\subsection{Collaborative Scheme of MPAI}
This work was accomplished in the MPAI-EEV group. Within this group, experts around the globe regularly gather and review the progress, and plan new efforts. 
MPAI-EEV plans on being an asset for AI-based E2E video coding by continuing to contribute novel tools and advanced models for the E2E video coding field, realizing the technological revolution in compression society. This research, contributed by MPAI-EEV experts, has constructed a systematic and comprehensive overview of what EEV has achieved in compressing UAV videos and facilitates future research works for related topics.

\subsection{Open Discussions}
With the AVS3, H.266/VVC, AV1 standards and their extension published in the past several years, a new set of video coding tools for next-generation video standards are under development in both standard research groups and academic community, which indicates that we are actually stepping into another cycle of finding higher coding efficiency evidence beyond the existing VVC standard. The E2E data-trained NN-based video coding opens a novel direction for coding efficiency improvement. Obviously, the NN-equipped video codec has such potential to bring us into a new stage. Moreover, the E2E optimization video coding is also capable of overcoming the problem of local optimization in hybrid coding frameworks. The learning objective is also feasible and tunable for both human perception and machine vision analysis. However, before we dive into neural video coding. There are a few preconditions that need to be carefully considered and resolved. 
\begin{itemize}
    \item {\bf Computational complexity reduction.} Reducing the computational complexity of neural networks is crucial, as networks are often over-parameterized and weights are sparsely distributed. The less salient neurons could be removed.  
    \item {\bf How to standardize} such data-trained codecs with millions of parameters is a critical issue if NN-based solutions are to be utilized. To enable intelligent-video-coding-compliant terminals and systems to decode latent representations without ambiguity, it is necessary to standardize them by defining the appropriate rules and assigning them to syntax elements. At the system level, NN descriptions, and latent representations should be parsed correctly by specific decoders. 
    \item The {\bf generalization and interpretation} of deep neural networks are of vital importance not only for the deep learning community but also for the video coding community. Robustness and solidness should always be the first priority for designing video coding algorithms.
    \item {\bf Data security-related issues} should be considered. Latent representations are derived from networks involving signal information that can be used to reconstruct the entire video stream. Such representations, however, are not encrypted and therefore exist the risk of sensitive information leakage. As such, trustworthy and robust coding network design plays a central role in real-world applications.
\end{itemize}

\subsection{Envision the Future}
In the foreseeable future, user requirements for higher-quality visual experiences are still increasing with more powerful hardware support. For entertainment, immersive media and meta-version scenarios are calling for techniques to handle multiple types of visual data, such as holograms, panorama videos, graphical data, and so on. 
To improve the ability to support downstream tasks using the neural-codec-compressed data, it is important to develop the learned codec to generate latent representations that are highly interpretable. 

By using such representations, it becomes possible to apply interactive coding techniques, which can enable a range of novel applications such as content editing and immersive interaction. This opens up new opportunities for compression-based approaches to provide versatile features and functionalities beyond traditional video compression methods.
Furthermore, active efforts to harmonize the learned visual data coding standard with other media data standards in emerging applications (such as short video in mobile devices and immersive media) will facilitate and expedite adoption in practical domains.

\FloatBarrier

\section{Conclusion}\label{conclude}
In recent years, AI-based video compression solutions have provided a comprehensive way to represent visual media compactly while describing intrinsic semantics. This has the potential to revolutionize current and future multimedia coding applications. This paper provides a systematic, comprehensive, and up-to-date report on the current progress and status of E2E-optimized neural approaches in video coding and standardization efforts. We adopt the MPAI-EEV as a learned video codec to showcase the progress. We believe that data-driven inspired deep models will be the dominant technologies and prosperous direction for future video compression, with capabilities beyond the current state-of-the-art hybrid framework.

\section*{Acknowledgement}
The authors would like to thank the guest editors and anonymous reviewers for their constructive comments to improve the quality of this manuscript. The experts of the MPAI-EEV group are also acknowledged for their contributions.
 
\bibliography{ref}
\bibliographystyle{IEEEtran}

\end{document}